\documentclass[pre,eqsecnum,twocolumn]{revtex4}
\usepackage[centertags]{amsmath}
\usepackage{amssymb}
\usepackage{amsthm}
\usepackage{graphicx}
\usepackage{epsfig}

\begin{document}

\title{Modelling light-driven proton pumps \\
in artificial photosynthetic reaction centers}

\author{Pulak Kumar Ghosh$^{1}$, Anatoly Yu. Smirnov$^{1,2}$, and Franco Nori$^{1,2}$}


\affiliation{$^1$ Advanced Science  Institute, The Institute of Physical and
Chemical Research (RIKEN), Wako-shi, Saitama, 351-0198, Japan  \\
 $^2$ Physics Department, Center for Theoretical Physics,  The University of Michigan,
  Ann Arbor, MI 48109-1040, USA }

\begin{abstract}
We study a model of a light-induced proton pump in artificial
reaction  centers. The model contains a molecular triad with four
electron states (i.e., one donor state, two photosensitive group
states, and one acceptor state) as well as a molecular shuttle
having one electron and one proton-binding sites. The shuttle
diffuses between the sides of the membrane and translocates protons
energetically uphill: from the negative side to the positive side
of the membrane, harnessing for this purpose the energy of the
electron-charge-separation produced by light. Using methods of
quantum transport theory we calculate the range of light intensity
and transmembrane potentials that maximize both the light-induced
proton current and the energy transduction efficiency. We also
study the effect of temperature on proton pumping.  The
light-induced proton pump in our model gives a quantum yield of
proton translocation of about 55\%. Thus, our results explain
previous experiments on these artificial photosynthetic reaction
centers.
\end{abstract}


\maketitle

\section{Introduction}
It would be desirable to create an artificial system that exploits the basic principles of natural photosynthesis in order to
produce energy in an usable form \cite{GustSc89,Gust01,Crabtree07,LaVan06,Wasiel06,Hambourger09,Barber09}. Indeed, natural
photosynthetic structures efficiently convert the energy of light into chemical form \cite{Barber09,Alberts}.

 The overall energy transduction process in  plant photosynthesis
 occurs through a number of strongly coupled successive stages (see, e.g.,
\cite{Alberts,GustSc89,Gust01}). In the first step, light of the appropriate wavelength is absorbed by a light harvesting complex.
The second step involves the conversion of electronic excitation energy to redox-potential in the form of the long-lived
transmembrane charge separation via multi-step electron transfer processes. The first two steps involve three constituents: (a)
light-absorbing pigments, (b) an electron acceptor, and (c) an electron donor.  In the third step, the energy stored in the electron
subsystem is used for energetically uphill proton pumping, which generates the proton motive force across the membrane.
\begin{figure}[htp]
\centering\includegraphics[width=8cm,angle=0,clip]{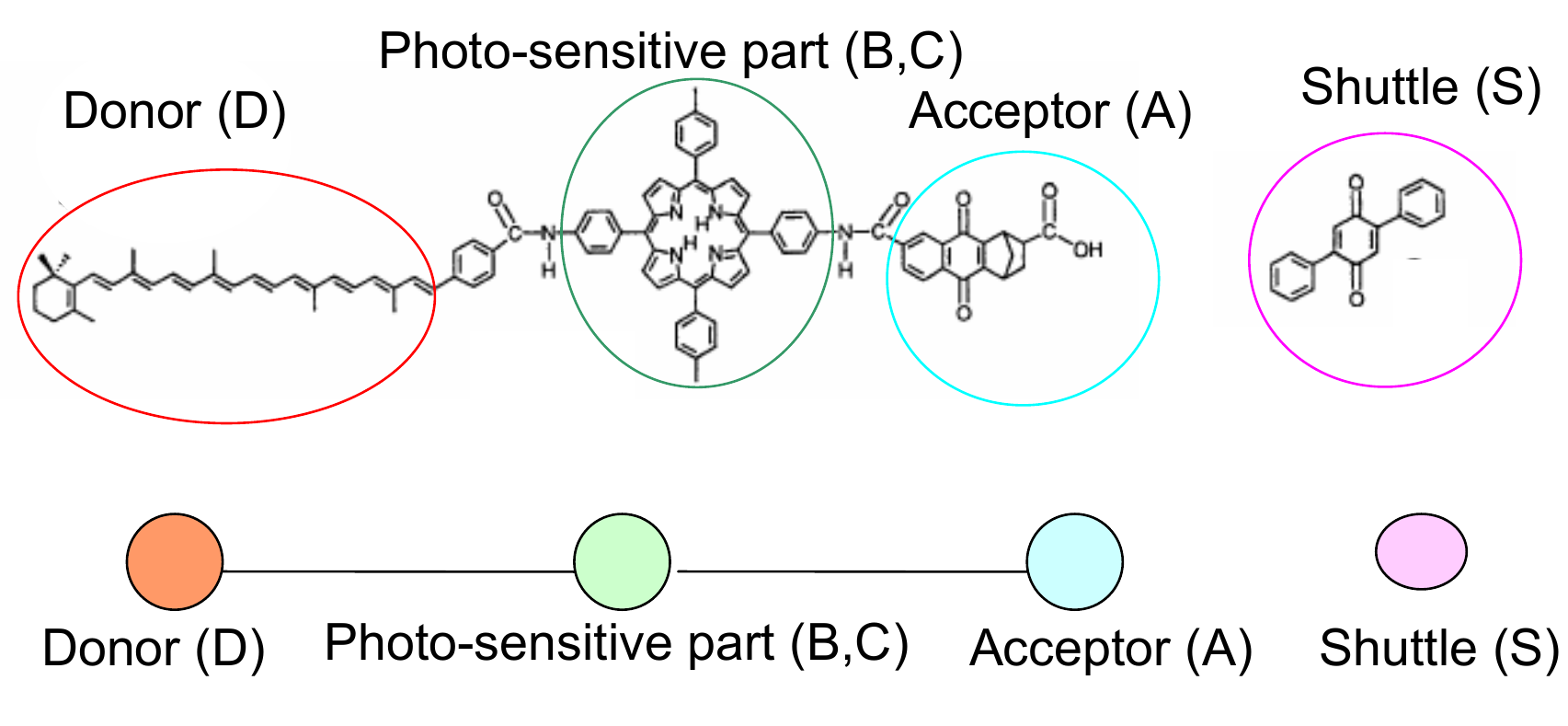}
\caption[]{  (Color online) The top figure presents the triad
(donor ``D", photo-sensitive part ``B,C", and acceptor ``A") and
the shuttle ``S" \cite{gali1,gali2}. These are enclosed by color
circles, which are schematically shown in the bottom figure. The
tetraarylporphyrin group acts as a photosensitive moiety (B,C)
(inside the green circle in the top structure). This is connected
to both a naphthoquinone moiety fused to a norbornene system with a
carboxylic acid group (which acts as an electron acceptor (A)) and
to a carotenoid polyene (which acts as an electron donor (D)).
2,5-diphenylbenzoquinone is the proton shuttle (S), denoted by a
pink hollow circle in the structure and by a solid pink circle in
the cartoon.}
\end{figure}

\begin{figure}[htp]
\centering\includegraphics[width=8cm,angle=0,clip]{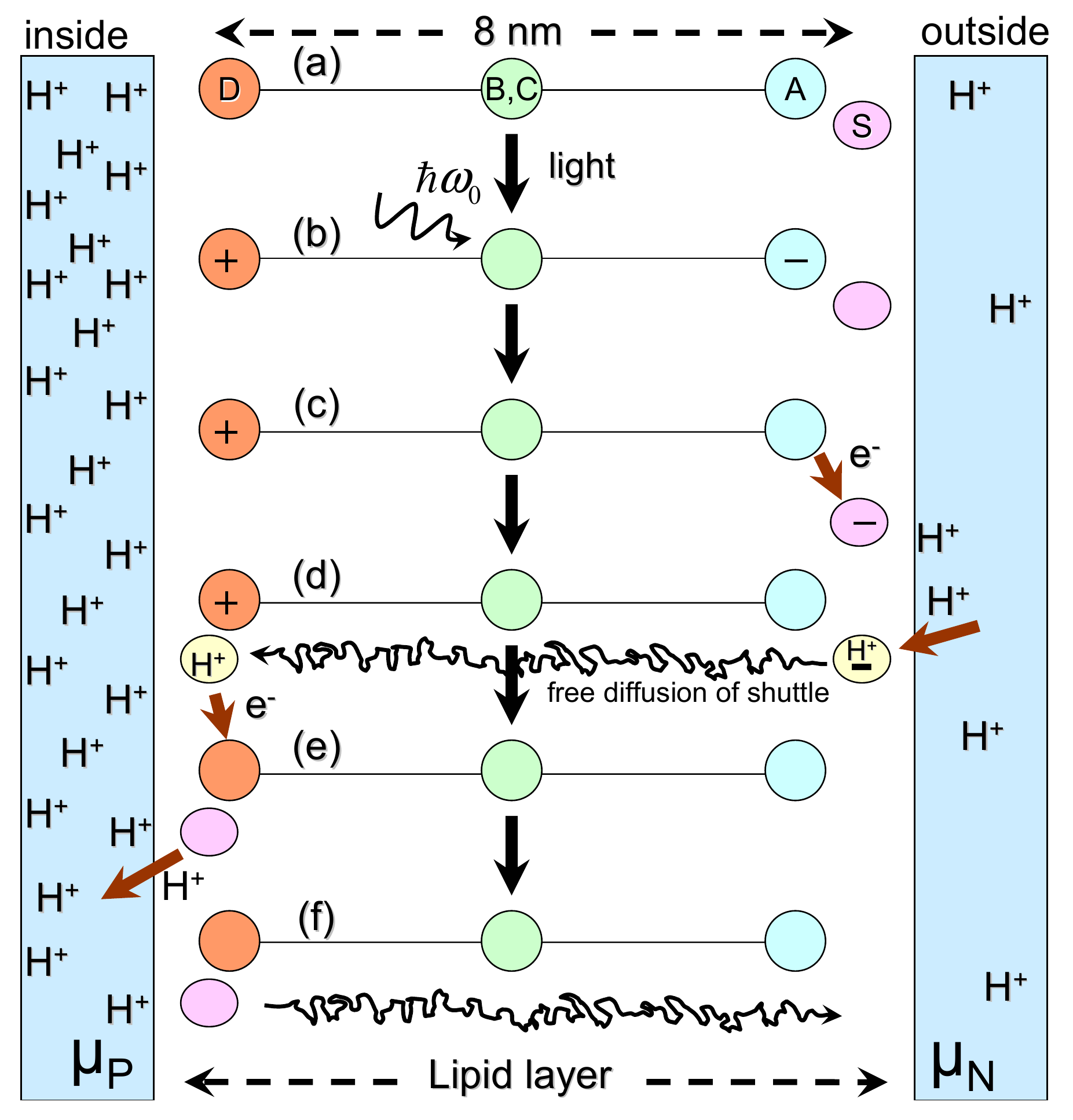}
\caption[]{(Color online) Schematic diagram of the light-induced
proton pump across the lipid bilayer in a liposomic membrane. A
molecular triad D--BC--C is symmetrically inserted in the lipid
bilayer. The different stages in  the proton pumping process are
here denoted by (a,b,c,d,e,f). The two bluish vertical rectangles
on both sides schematically represent two proton reservoirs with
electrochemical potentials $\mu_{\rm{P}}$ and $\mu_{\rm{N}}$. These
two proton reservoirs correspond to the aqueous phases inside and
outside of the liposome, respectively. The shuttle molecule S, is
shown as a pink-colored oval and the protonated neutral shuttle is
shown as a yellow oval. This shuttle freely diffuses in (d) (the
black scribbled curves represent the thermal stochastic motion of
the shuttle) across the membrane to transport a proton from the
lower proton potential $\mu_{\rm{N}}$ to the higher proton
potential $\mu_{\rm{P}}$ side of the membrane, where
$(\mu_{\rm{P}}-\mu_{\rm{N}})$ denotes the total potential
difference between the two reservoirs. }
\end{figure}

The study of natural photosynthesis has inspired researchers to  perform the photo-induced energy transduction processes in the
laboratory \cite{GustSc89,Gust01,gali1,gali2,Crabtree07,LaVan06,Wasiel06,Hambourger09,Barber09,moo1,Imahori04}. A convenient
approach to photosynthesis in artificial reaction centers is to use synthetic pigments, electron acceptors and electron donors that
are very similar in molecular structure to natural pigments (e.g., chlorophylls, carotenoids and quinones). In this direction, the
experimental model proposed in Refs.~\cite{gali1,gali2} provides a paradigm for the conversion of light energy to a proton potential
gradient. These seminal works \cite{gali1,gali2} have motivated research in the design and synthesis of new artificial
photosynthetic systems \cite{bho1,pal12,pol1} (i.e., light-harvesting antennas and reaction centers) and triggered considerable
experimental \cite{syk2,ima1,sah1,riz1,ima2} and theoretical \cite{sod1,cri2,oka1,Andreas} activities to investigate more
sophisticated and more efficient mechanisms for the conversion of light energy.

The transformation of light energy into the electrochemical gradient of protons across the membrane can be quantitatively
characterized by the quantum yield (or quantum efficiency), $\Phi$, of proton translocation. This parameter is defined as the total
number of translocated protons divided by the number of photons absorbed by the triad \cite{gali1}. A quantum yield of the order of
0.4\% has been measured in Ref.~\cite{gali1}. A much higher quantum efficiency, $\Phi \sim 7\%,$ for the conversion of photons into
ATP molecules, was found in Ref.~\cite{gali2}. As argued in Ref.~\cite{gali2}, the actual quantum yield of ATP formation could be of
the order of 15\%, if we take into account the real rate of light absorbance, which is $\sim$~50\%. Near four protons are necessary
for the synthesis of a single ATP molecule. This means that the real quantum yield $\Phi$ of proton translocation measured in
Ref.~\cite{gali2} can be about 60\%. The total thermodynamic (or power-conversion) efficiency, $\eta $, of the light-to-ATP
conversion process is estimated in Ref.~\cite{gali2} as $\eta \sim$~4\%.

In the present paper, using methods from quantum transport theory \cite{Wingr93,PumpPRE08,FlagPRE08,PumpTime08}, we analyze the
photoinduced electron and proton transfer in a molecular triad inserted into a liposomal membrane, which contains a single molecular
shuttle. We calculate the photon-to-proton quantum yield $\Phi \sim $ 55\% (and the thermodynamic efficiency $\eta \sim$ 6.3\%) for
the resonant tunneling conditions, when the reorganization energy, $\lambda$, of the electron transitions matches the detuning
$\delta$ between the electron energy levels: $\lambda \sim \delta.$

We note that due to a small optimal value of the  reorganization
energy ($\lambda \sim 400$~meV) the charged recombination process
in the triad is described by the inverted region of the Marcus
formula \cite{rd,bath2,Imahori02}. This further enhances the
performance of the system. Our results explain experiments made in
Ref.~\cite{gali2} using artificial photosynthetic centers. The
obtained power-conversion efficiency corresponds to the highest
value, $\eta \sim$ 6.5\%, achieved recently with polymer solar
cells \cite{KimSc07}. It is expected that the proton current and
the efficiency should increase with increasing the number of the
shuttles in the membrane.

This article is organized as follows. In Sec. II (see also the
Appendix) we introduce the basis set for the system and write the
Hamiltonian of the problem. In Sec. III, we present the master
equation for the density matrix coupled to the Langevin equation
describing the diffusive motion of the shuttle in the lipid
bilayer. In Sec. IV, we numerically solve these equations and
analyze the light-induced proton pumping process. In Sec. V we
summarize our results.

\section{Model}
We use a slightly modified version of the well-accepted model already presented, e.g., in Refs.~\cite{gali1,gali2}. In this model
 the reaction center is a molecular triad containing an electron donor and an electron acceptor both
linked to a photosensitive porphyrin group (shown in Fig.~1). The triad molecule (D--BC--A) is inside the bilayer of a liposome. The
lipid bilayer also contains freely diffusing 2,5 diphenylbenzoquinones, acting as proton shuttles. The molecular triad absorbing a
photon establishes a negative charge near the outer surface and a positive charge near the inner surface of the liposome, by
generating charge separated species D$^+$--BC--A$^-$. The freely diffusing quinone shuttle translocates an electron-proton pair
across the membrane and neutralizes the molecular triads.
\begin{figure}[htp]
\centering\includegraphics[width=8cm,angle=0,clip]{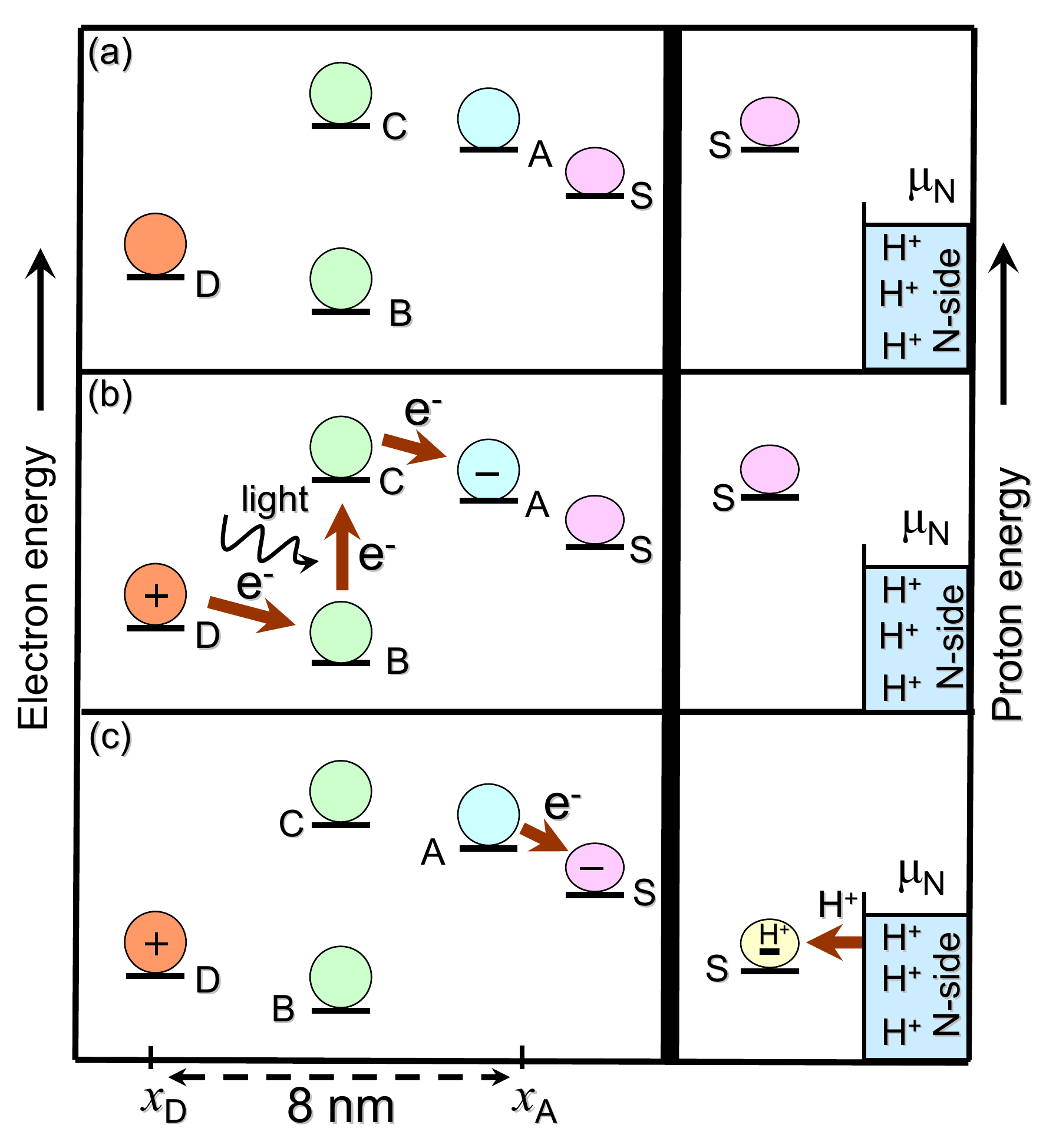}
\caption[]{ (Color online) Energy diagram depicting the energy
levels of states involved in an artificial photosynthetic reaction
center, \textit{before} the diffusion of the shuttle to the
P-reservoir. The subfigures (a,b,c) correspond to the stages
(a,b,c) in Fig.~2. The left and right panels represent electron and
proton energy levels, respectively. The abbreviations D, B, C, A, S
are the same as used in the text and in Fig.~1. Also, $x_{\rm{D}}$
and $x_{\rm{ A}}$ represent the spatial coordinates of the sites D
and A, respectively. The thick brown arrows denote the path the
electrons follow in this energy diagram, generating charge
separation, in (b), and shuttle charging and protonation in (c).
Initially, light excites an electron from B to C, and eventually to
A, making it A$^-$. Afterwards, in (b), the donor D loses an
electron, thus becoming D$^+$, and that electron moves to BC. Later
on, the shuttle S in (c) receives the electron from A. }
\end{figure}
\begin{figure}[htp]
\centering\includegraphics[width=8cm,angle=0,clip]{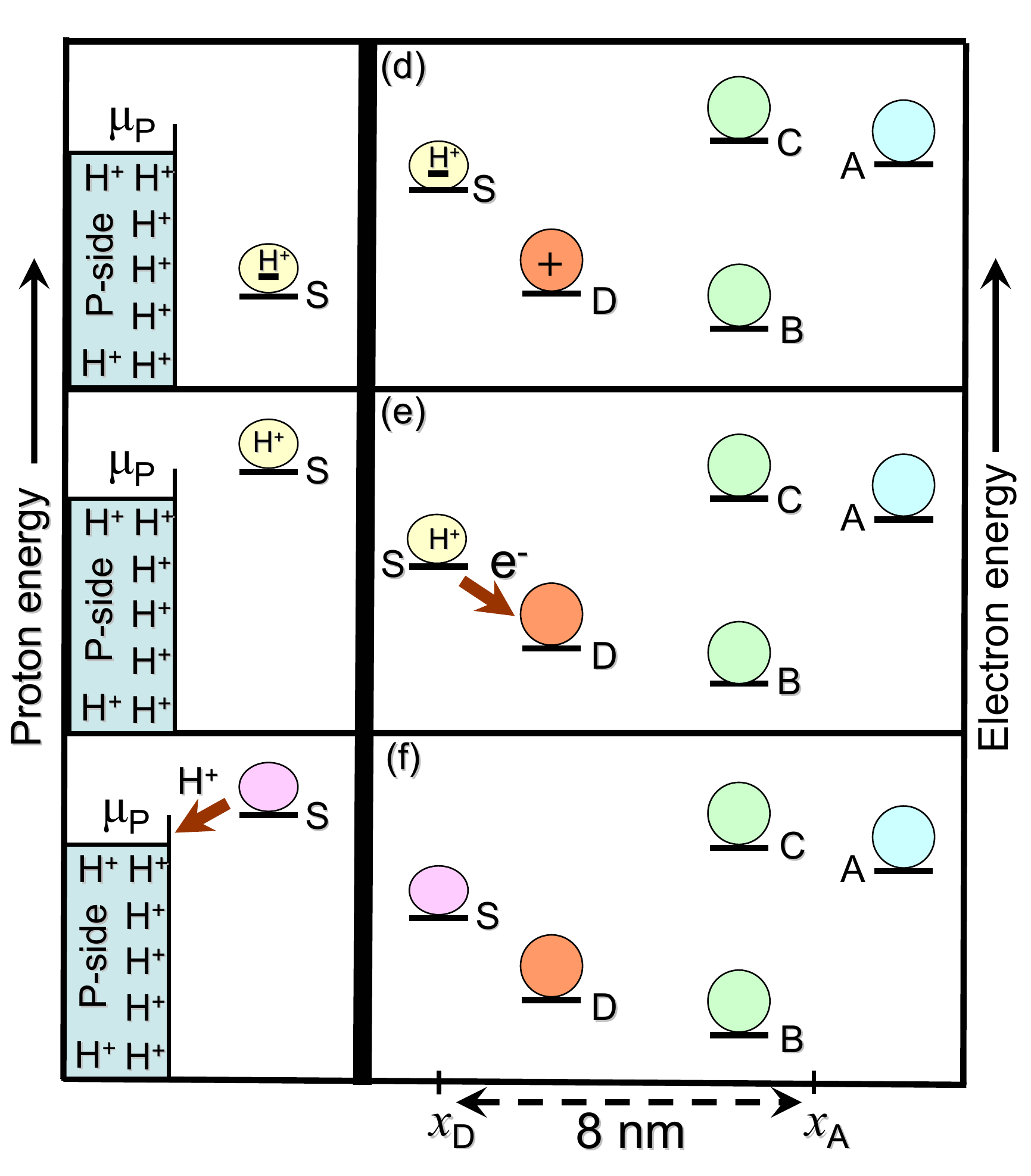}
\caption[]{  (Color online) Energy levels involved in an artificial
photosynthetic reaction center. This figure is similar to Fig.~3,
but now the energy profile corresponds to the stage \emph{after}
the shuttle diffuses to the P-reservoir. Here the subfigures
(d,e,f) correspond to the stages (d,e,f) in Fig.~2. The left and
right panels represent proton and electron energy levels,
respectively. The thick brown arrows denote the path followed by
the electron (e) and proton (f). In (d), an electron on the shuttle
S moves to the donor site D, neutralizing it in (e). This electron
transition in the right panels increase the proton energy of the
shuttle, as shown in the left panels (from (d) to (e)). The proton
finally leaves the shuttle in the left panel of (f). }
\end{figure}
 In Fig.~2 we schematically illustrate the process of light-induced proton
pumping in liposomes by artificial photosynthetic reaction centers \cite{gali1,gali2}. The transmembrane proton pumping requires a
symmetric arrangement of the molecular triad (of length $\sim 8$ nm) inside the bilayer and with a specific direction: with the
acceptor (A) site towards the outer membrane of the liposome (the negative (N) side of the membrane), and with the donor (D) towards
the inside of the liposome (the positive (P) side of the membrane) \cite{gali1,gali2}.

The energy diagrams of the electron and proton sites are shown in Figs.~3 and 4. There are two electrons in the system, one of which
is initially on the D site, and another electron is on the lower energy level B. The quinone molecular shuttle has one electron
state S (denoted by S, instead of S$_{\rm{e}}$), and one proton state Q (denoted here by Q instead of S$_{\rm{p}}$). Thus, S denotes
the shuttle electron state and Q denotes the shuttle proton state.

The overall process leading to the proton translocation from the N-reservoir with a lower proton potential, $\mu_{\rm{N}}$, to the
P-reservoir with a \emph{higher} electrochemical potential, $\mu_{\rm{P}}$, can be considered as a sequence of eight stages (most of
which are shown in Fig.~2).
\begin{itemize}
\item Step I: The photosensitive moiety of the molecular triad absorbs light and an electron goes from the ground state B to the
excited state C (see Fig.~3b). \item Step II: The unstable excited state C transfers the electron to the acceptor A, producing an
unstable charge-separated intermediate species D--BC$^{+}$--A$^{-}$. \item Step III: The unstable intermediate charge-separated
species is rapidly rearranged to a relatively stable charge-separated form (D$^{+}$--BC--A$^{-}$) by the thermal electron transfer
from the state D to the state B$^+$ having a lower energy than the state D (Fig.~3b). \item Step IV: The shuttle in the position
near the N-side of the membrane accepts an electron from A$^{-}$ and becomes negatively charged. \item Step V: The shuttle molecule
receives a proton from the N-reservoir and becomes neutralized (Fig.~3c right panel). \item Step VI: The neutral shuttle slowly
diffuses through the lipid bilayer and carries the electron and the proton to the P-side of the membrane and to the D-site (stage
(d) in Fig.~2). \item Step VII: The shuttle gives away the electron to the positively charged site D$^{+}$ (stage (e) in Fig.~2 and
Fig.~4e). \item Step VIII: The shuttle is deprotonated by donating the proton to the P-reservoir (Fig.~4f).
\end{itemize}
This sequence of eight steps describes the photo-induced electron transfer that generates the intra-membrane redox potential, which
in turn drives the energetically uphill vectorial translocation of protons by the shuttle.

Electrons in the states $i$ (= D,B,C,A,S) and protons in the state Q are characterized by the corresponding Fermi operators
$a_i^+,a_i$ and $b_{\rm{Q}}^+,b_{\rm{Q}}$, with the electron population operator $n_i$ and the proton population $n_{\rm{Q}}$. We
assume that each electron or proton state can be occupied by a single electron or a single proton. Spin degrees of freedom are
neglected. The proton site on the shuttle, denoted by Q, can be populated from the N-reservoir provided the shuttle is within the
transition length $L_{\rm{Q}}$ from the N-side of the membrane. The protonated shuttle, located within the transition (or tunneling)
range from the P-side of the membrane, can donate its proton to the P-reservoir. Protons in the reservoirs are described by the
Fermi operators $d_{k\alpha}^+,d_{k,\alpha}$, where $\alpha = \rm {N,P} $; and $k$ is an additional parameter which has the meaning
of a wave vector in condensed matter physics \cite{Wingr93,PumpPRE08,FlagPRE08,PumpTime08}. The number of protons in the reservoirs
is determined by the operator $\sum _{k}N_{k\alpha}$, with $N_{k\alpha}=d_{k\alpha}^+ d_{k\alpha}$.

\subsection{Hamiltonian}
The Hamiltonian of the electron-proton system,
\begin{equation}
H = H_{0} + H_{\rm {dir}} + H_{\rm{tr}} + H_{\rm{B}},
\end{equation}
 has a term $H_{0}$ related to the energies $E_i$ of the electron eigenstates
($i$ = D,B,C,A,S), and to the energy $\epsilon_{\rm{Q}}$ of a proton, on the shuttle:
\begin{eqnarray}
H_{0} &=& \sum _{i} E_i n_i + \epsilon_{\rm{Q}} n_{\rm{Q}} +
u_{\rm{DB}} (1-n_{\rm {D}})(1-n_{\rm {B}}-n_{\rm{C}})  \nonumber\\
&-& u_{\rm {DA}} (1-n_{\rm {D}})n_{\rm{A}} -
u_{\rm{BA}}(1-n_{\rm{B}} -n_{\rm{C}})n_{\rm{A}} \nonumber\\
&-&u_{\rm{SQ}}\; n_{\rm{S}} \;n_{\rm{Q}}.
\end{eqnarray}
We include here the electrostatic interaction  between the electron sites, $u_{\rm{DB}},u_{\rm{DA}},u_{\rm{BA}}$, and the Coulomb
attraction $u_{\rm{SQ}}$ between the electron and proton sites on the shuttle. It is assumed that the empty donor state D (with
$n_{\rm{D}}=0$) as well as the empty photosensitive group B and C ($n_{\rm {B}} + n_{\rm{C}} = 0$) have positive charges, and
$u_{\rm {DB}}=u_{\rm {DC}},\, u_{\rm{CA}}=u_{\rm{BA}}.$

The term,
\begin{eqnarray}
H_{\rm{dir}} &=&-\Delta_{\rm{DB}}\; a_{\rm{D}}^{\dag}a_{\rm{B}} -\Delta_{\rm{AC}}\; a_{\rm{A}}^{\dag} a_{\rm{C}} - \Delta_{\rm{DS}}
(x) \; a_{\rm{D}}^
{\dag} a_{\rm{S}}\nonumber \\
&-& \Delta_{\rm{AS}} (x)\; a_{\rm{A}}^{\dag} a_{\rm{S}} - F(t) \; a_{\rm{B}}^{\dag} a_{\rm{C}} + h.c.,
\end{eqnarray}
describes the tunneling of electrons between the  sites D--B, C--A, A--S, and D--S, with the corresponding amplitudes
$\Delta_{ii'}$. Notice that the tunneling elements $\Delta_{\rm{DS}}(x)$ and $\Delta_{\rm{AS}}(x)$ depend on the shuttle position
$x$. The Hamiltonian $H_{\rm{dir}}$ is also responsible for the electron transitions between the states B and C  induced by the
electromagnetic field (light), $F(t)= F_0 \exp(i\omega_0 t)$, with a frequency $\omega_0$ and an amplitude $F_0$. Proton transitions
between the shuttle (site Q) and the N- and P-proton reservoirs are governed by the Hamiltonian
\begin{eqnarray}
H_{\rm{tr}} =-\sum _{k\alpha}T_{k\alpha}(x)\;d_{k\alpha}^{\dag}b_{\rm{Q}}  - \sum _{k\alpha}
T_{k\alpha}^*(x)\;b_{\rm{Q}}^{\dag}d_{k\alpha},
\end{eqnarray}
with the position-dependent coefficients, $T_{k\alpha}(x)$.  We have chosen the following form of $T_{k\alpha}(x)$:
\begin{eqnarray}
 T_{kN}(x) &=&
T_{kN} \theta[x -(x_N-L_Q)],\nonumber\\ T_{kP}(x) &=& T_{kP}
\theta[x_P+L_Q - x],\nonumber
\end{eqnarray}
 where $\theta(x)$ is the Heaviside step function, and the
parameter $L_{\rm Q}$ defines the proton loading range of the shuttle.

\subsection{Interaction with the environment}
To take into consideration the effect of a dissipative environment we consider the well-known system-reservoir model
\cite{rd,bath2,bath1}, where the medium surrounding the active sites is represented by a system of harmonic oscillators with the
Hamiltonian:
\begin{eqnarray}
H_{\rm{B}}=\sum _{j}\left[ \frac{p_j^2}{2m_j}+ \frac{m_j\omega_j^2
}{2} \left( x_j +\frac{1}{2} \sum_i x_{ji}n_i\right)^2 \right],
\end{eqnarray}
where ${x_j,p_j}$ are the positions and momenta  of the oscillators with effective masses $m_j$ and frequencies $\omega_j$. The
parameters $x_{ji}$ determine the strengths of the coupling between the electron subsystem and the environment. The system of
independent oscillators are conveniently characterized by the spectral functions $J_{ii'}(\omega)$, defined by
\begin{eqnarray}
J_{ii'} (\omega) = \sum_j \frac{m_j\omega_j^3 (x_{ji}-x_{ji'})^2}{2} \delta(\omega -  \omega _ j),
\end{eqnarray}
so that the reorganization energy $\lambda_{ii'}$,  related to the $i \rightarrow i'$ transition,  has the form
\begin{eqnarray}
\lambda_{ii'} = \int_{0}^{\infty} \frac{d \omega} {\omega} J_{ii'}(\omega)
 = \sum_j \frac{m_j\omega_j^2 (x_{ji}-x_{ji'})^2}{2}.
\end{eqnarray}
With the unitary transformation $\hat{U}=\prod _i \hat{U}_i$, where
\begin{eqnarray}
\hat{U}_i = \exp{\left[\frac{i}{2} \sum_{j} p_j x_{ji}n_i\right]},
\end{eqnarray}
we can transform the Hamiltonian $H$ to the form $H'=U^{\dag}HU$, becoming (after dropping the prime)
\begin{eqnarray}
H &=& H_{0} - \sum_{ii'} \Delta_{i i'}\; e^{(i/2)  (\xi_{i}
-\xi_i')} \;a^{\dag}_{i'} \;a_i \nonumber\\&-&  F(t)
e^{-(i/2)(\xi_{\rm{B}} -\xi_{\rm{C}})} \;a^{\dag}_{\rm{B}}
a_{\rm{C}}
 -F^*(t) \;a^{\dag}_{\rm{C}} \;a_{\rm{B}}\; e^{(i/2)(\xi_{\rm{B}} -\xi_{\rm{C}})} \nonumber\\
 &-& \sum_{k\alpha} T_{k\alpha}(x)\;d_{k\alpha}^ {\dag}\; b_{\rm{Q}} -
 \sum_{k\alpha} T_{k\alpha}^*(x)\;b_{\rm{Q}}^{\dag} \;d_{k\alpha}\nonumber\\&+&
  \sum _{j}\left(\frac{p_j^2}{2m_j}+\frac{m_j\omega_j^2 x_j^2}
 {2}\right),
\end{eqnarray}
where $\alpha$ = N,P, and the tunneling coefficients, $\Delta_{i i'}^* =  \Delta_{i' i}$, take non-zero values only for transitions
between the sites D and B, A and C, A and S, as well as D and S.
 The stochastic phase operator $\xi_i$ is given by
\begin{eqnarray}
\xi_i = \frac{1}{\hbar} \sum_{j} p_j x_{ji}.
\end{eqnarray}
The result of this transformation follows from the fact that,  for an arbitrary function $\Phi(x_j)$, the operator $\hat{U}$
produces a shift of the oscillator positions:
\begin{eqnarray}
\hat{U}^{\dag}\Phi(x_j)\hat{U}=\Phi \left(x_j + \frac{1}{2}\sum_i x_{ji}n_i\right).
\end{eqnarray}
This transformation also results in the phase factors for the electron
 amplitudes (see Eq.~(9)).

 The basis sets, composed of the electron-proton eigenstates, and
 their corresponding energy eigenvalues are presented in an
 Appendix. Thus, the reader is encouraged to read this short
 Appendix before proceeding further.

\section{Time evolution of density matrix}

\subsection{Master equations}
 To describe the time evolution of the diagonal elements
  of the density matrix, $\langle \rho_m\rangle$,  we
write  the Heisenberg equation for the operators $\rho_m$ with the subsequent averaging over the environment fluctuations and over
the states of the proton reservoirs:
\begin{eqnarray}
\langle \dot{\rho}_m \rangle = - \langle i [\rho_m, H_{\rm dir}]_{-}\rangle  - \langle i [\rho_m, H_{\rm tr}]_{-}\rangle.
\end{eqnarray}

The protons in the reservoirs ($\alpha$ = N,P) are characterized by the Fermi distributions,
\begin{eqnarray}
F_{\alpha}(E_{k\alpha})= \left[\exp\left(\frac{ E_{k\alpha}-\mu_{\alpha}}{T}\right)
 +1 \right]^{-1}.
\end{eqnarray}
with the temperature $T$ ($k_{\rm{B}} = 1$). The electrochemical potentials $\mu_{\rm{N}}$ and $\mu_{\rm{P}}$, correspond to the
negative (N) and positive (P) proton reservoirs, respectively. The proton motive force ($\Delta \mu$) across the membrane is given
by
\begin{eqnarray}
\Delta \mu = \mu_{\rm{P}}-\mu_{\rm{N}} =  V - \frac{2.3 \ R T}{F}\left(\Delta pH\right), \label{dMu}
\end{eqnarray}
where $R$ and $F$ are the gas constant and Faraday constant, respectively, and $V$ is the transmembrane voltage gradient. Hereafter
we change $\Delta \mu$ by changing the $pH$ of the solution by $\Delta pH$.

The contribution of the transitions between  the shuttle and the proton reservoirs to the time evolution of the density matrix is
described by the second term in the right hand side of Eq.~(12), which can be calculated with methods of quantum transport theory
\cite{Wingr93,PumpPRE08}
\begin{eqnarray}
\langle i [\rho_m, H_{\rm tr}]_{-}\rangle =   \sum _{n} \left[ \gamma_{nm}^{\rm tr}(x)\langle \rho_m \rangle - \gamma_{mn}^{\rm
tr}(x)\langle \rho_n \rangle  \right],
\end{eqnarray}
with the relaxation matrix
\begin{eqnarray}
\gamma_{mn}^{\rm tr}(x) &=& \sum_{\alpha} \Gamma_{\alpha}(x)
\left\{ |b_{Q,mn}|^2[1-F_{\alpha}(\omega_{nm})]\right. \nonumber\\
&+& \left. |b_{Q,nm}|^2 F_{\alpha}(\omega_{mn})\right\}.
\end{eqnarray}
Here we introduce the frequency-independent coefficients,
\begin{eqnarray}
\Gamma_{\alpha}(x) = 2\pi \sum_{k} |T_{k\alpha}(x)|^2 \;\delta (\omega -E_{k\alpha}),
\end{eqnarray}
which determine the transition rates between the shuttle state Q and the sides of the membrane (N- and P-reservoirs). Notice that
these coefficients are functions of the shuttle position $x$.

The transitions between the electron levels are described by the Hamiltonian $H_{\rm dir}$, which can be written as
\begin{eqnarray}
H_{\rm dir} = - \sum_{mn} {\cal A}_{mn} \;\rho_{m,n} - \sum_{mn} \rho_{n,m} \;{\cal A}_{mn}^{\dag},
\end{eqnarray}
with the functions
\begin{eqnarray}
{\cal A}_{mn} &=& Q_{\rm{DB}} (a_{\rm{B}}^{\dag}a_{\rm{D}})_{mn}  +
Q_{\rm{CA}} (a_{\rm{A}}^{\dag}a_{\rm{C}})_{mn} + Q_{\rm{SA}}
(a_{\rm{A}}^{\dag}a_{\rm{S}})_{mn}
  \nonumber\\
&+& Q_{\rm{SD}} (a_{\rm{D}}^{\dag}a_{\rm{S}})_{mn} + Q_{\rm{CB}} (a_{\rm{B}}^{\dag}a_{\rm{C}})_{mn},
\end{eqnarray}
which are defined as superpositions of the heat-bath operators
\begin{eqnarray}
Q_{i i'} &=& \Delta_{i' i} \exp[(i/2)(\xi_{i} - \xi_{i'})]
\nonumber\\
&=& \Delta_{i' i} \exp[(i/2)\sum_j p_j(t)(x_{j i} - x_{ji'})],
\end{eqnarray}
for the pairs of the electron sites $(i i')$ = (DB),(CA),(SA),(SD),
 whereas for the pair (CB) we have
\begin{equation}
Q_{\rm{CB}} = F_0 \exp(i\omega_0 t) \exp[(i/2)\sum_j p_j(t)(x_{j C} - x_{j B})],
\end{equation}
In the case of a high-enough temperature of the bath \cite{bath2}, the cumulant functions of the unperturbed operators $Q_{i
i'}^{(0)}$ are determined by the relations:
\begin{eqnarray}
\langle Q_{i i'}^{(0)}(t),Q_{i i'}^{(0)\dag}(t')\rangle &=&
|\Delta_{i' i}|^2 e^{- i \lambda_{i i'}(t-t')} e^{-\lambda_{i i'} T
(t-t')^2},
\nonumber\\
\langle Q_{i i'}^{(0)\dag}(t),Q_{i i'}^{(0)}(t)\rangle &=&
|\Delta_{i' i}|^2 e^{ i \lambda_{i i'}(t-t')} e^{-\lambda_{i i'} T
(t-t')^2}.\nonumber\\
\end{eqnarray}
The contribution of the electron transitions to Eq.~(12) is determined by the term
\begin{eqnarray}
\langle -i [\rho_m, H_{\rm dir}]_{-}\rangle =  i \sum_n \langle
{\cal A}_{mn} \rho_{mn} - {\cal A}_{nm} \rho_{nm} \rangle  + h.c.
\nonumber \\
 \end{eqnarray}
Within the theory of open quantum systems developed in
Refs.~\cite{PumpTime08}, the correlation  function $\langle {\cal
A}_{mn} \rho_{mn}\rangle $ is proportional to the density matrix
elements of the system, $\langle \rho_m \rangle,$ with coefficients
defined by the unperturbed correlators $\langle {\cal
A}_{mn}^{(0)}(t),{\cal A}_{mn}^{(0)\dag}(t')\rangle$ of the bath
operators:
\begin{eqnarray}
\langle {\cal A}_{mn}(t) \rho_{mn}(t) \rangle &=& i \int dt_1
\theta(t-t_1) e^{i\omega_{mn}(t-t_1)}\nonumber
\\
 &\times& \left\{
\langle {\cal A}_{mn}^{(0)}(t),{\cal A}_{mn}^{(0)\dag}(t_1)\rangle
\langle \rho_m (t) \rangle \right. \nonumber
\\
 &-& \left.\langle {\cal
A}_{mn}^{(0)\dag}(t_1),{\cal A}_{mn}^{(0)}(t)\rangle \langle \rho_n
(t) \rangle \right\},\nonumber\\
\end{eqnarray}
where
\begin{eqnarray}
\langle {\cal A}_{mn}^{(0)}(t),{\cal A}_{mn}^{(0)\dag}(t_1)\rangle
&=&  \langle Q_{\rm{CB}}^{(0)}(t),Q_{\rm{CB}}^{(0)\dag}(t_1)\rangle
|(a_{\rm{B}}^{\dag}a_{\rm{C}})_{mn}|^2  \nonumber
\\
&+& \langle Q_{\rm{DB}}^{(0)}(t),Q_{\rm{DB}}^{(0)\dag}(t_1)\rangle
|(a_{\rm{B}}^{\dag}a_{\rm{D}})_{mn}|^2 \nonumber
\\
&+& \langle Q_{\rm{CA}}^{(0)}(t),Q_{\rm{CA}}^{(0)\dag}(t_1)\rangle
|(a_{\rm{A}}^{\dag}a_{\rm{C}})_{mn}|^2 \nonumber
\\
&+& \langle Q_{\rm{SA}}^{(0)}(t),Q_{\rm{SA}}^{(0)\dag}(t_1)\rangle
|(a_{\rm{A}}^{\dag}a_{\rm{S}})_{mn}|^2 \nonumber
\\&+& \langle
Q_{\rm{SD}}^{(0)}(t),Q_{\rm{SD}}^{(0)\dag}(t_1)\rangle
|(a_{\rm{D}}^{\dag}a_{\rm{S}})_{mn}|^2,\nonumber\\
\end{eqnarray}

and the reverse expression can be obtained for the correlator $\langle {\cal A}_{mn}^{(0)\dag}(t_1),{\cal A}_{mn}^{(0)}(t)\rangle$.
The formula (24) is valid in the case of weak tunneling and  weak driving force $F_0$. The effects of quantum coherence are also
neglected here.

Finally, we derive the master equation for the density matrix of the system,
\begin{eqnarray}
\langle \dot{\rho}_m\rangle + \sum_{n}\gamma_{nm}(x) \langle \rho_m\rangle = \sum_{n}\gamma_{mn}(x) \langle \rho_n\rangle,
\end{eqnarray}
with the total relaxation matrix
\begin{eqnarray}
\gamma_{mn}(x) &=& \gamma_{mn}^{\rm tr}(x) +
(\kappa_{\rm{DB}})_{mn} +  (\kappa_{\rm{CA}})_{mn} \nonumber
\\ &+&
(\kappa_{\rm{SA}})_{mn} + (\kappa_{\rm{SD}})_{mn} +
(\kappa_{\rm{CB}})_{mn},
\end{eqnarray}
containing the contribution of proton transitions to and  from the
shuttle, $\gamma_{mn}^{\rm tr}(x)$, together with the Marcus rate
$(\kappa_{\rm{CB}})_{mn}$ describing the light-induced electron
transfer between the sites $B$ and $C$:
\begin{eqnarray}
(\kappa_{\rm{BC}})_{mn}&=&|F_0|^{2}\sqrt{\frac{\pi}{\lambda_{\rm{BC}}T}}
|(a_{\rm{B}}^{\dag}a_{\rm{C}})_{mn}|^2 \nonumber
\\
&\times&\exp\!\left[-\;\frac{\left( \omega_{mn}+\omega_0
+\lambda_{\rm{BC}}\right)^2}{4\lambda_{\rm{BC}}T}\right] \nonumber
\\
&+& |F_0|^{2}\sqrt{\frac{\pi}{\lambda_{\rm{BC}}T}}
|(a_{\rm{B}}^{\dag}a_{\rm{C}})_{nm}|^2 \nonumber
\\
&\times&\exp\!\left[-\;\frac{\left( \omega_{mn}-\omega_0
+\lambda_{\rm{BC}}\right)^2}{4\lambda_{\rm{BC}}T}\right],
\end{eqnarray}
as well as the rates related to the electron transfers between the pairs of sites $(i i')$ = (DB),(CA),(AS), and (DS):
\begin{eqnarray}
(\kappa_{i i'})_{mn}&=& |\Delta_{i' i}|^2
\sqrt{\frac{\pi}{\lambda_{i i'}T}} \left[
|(a_{i'}^{\dag}a_i)_{mn}|^2 + |(a_{i'}^{\dag}a_i)_{nm}|^2 \right]
\nonumber
\\
&\times&\exp\left[-\frac{\left( \omega_{mn} + \lambda_{i
i'}\right)^2}{4\lambda_{i i'}T}\right].
\end{eqnarray}
We note that the tunneling coefficients $\Delta_{\rm{AS}} $ and
$\Delta_{\rm{DS}}$ depend on the shuttle position $x$.

\subsection{Equation of motion for the shuttle}

We assume that the shuttle moves along the linear molecular triad (Fig.~1), and this motion can be described by the overdamped
Langevin equation for the shuttle position $x$:
\begin{eqnarray}\label{langevin}
\eta_{\rm{drag}} \;\frac{dx}{dt} = -\;\frac{dU(x)}{dx}+\zeta(t).
\end{eqnarray}
Here $\eta_{\rm{drag}}$ is the drag coefficient of the shuttle in the lipid membrane, and the thermal fluctuation of the medium is
modelled by a zero-mean delta-correlated Gaussian fluctuation force $\zeta(t)$, $\langle \zeta(t)\rangle = 0, $
\begin{eqnarray}
\langle \zeta(t)\zeta(t')\rangle = 2\eta_{\rm{drag}} T \delta(t-t'),
\end{eqnarray}
where $T$ is the temperature of the medium ($k_B$=1).  The diffusion of the shuttle is determined by the diffusion coefficient
$D_{\rm{s}}=T/\eta_{\rm{drag}}$. The potential $U(x)$ in Eq.~(\ref{langevin}) is responsible for the spatial confinement of the
hydrophobic shuttle (quinone) inside the lipid membrane.

\section{Results and discussions}

To analyze the light-induced proton pumping  process
quantitatively, we use the standard Heun's algorithm to numerically
solve the
 twenty coupled master equations (26) along with the equation (30)
for the shuttle. For initial conditions we have  assumed that at
$t=0$, $\rho_{1,1} = 1$, and the other elements of the density
matrix are zero (this corresponds to one electron on site D and
another electron on site B with no electrons and no protons on the
shuttle). We also assume that at $t=0$ the shuttle is located
nearby the acceptor (A): $x(t=0) = x_{\rm A} \simeq x_{\rm{N}}$.
Throughout our simulation we focus on the long-term asymptotic
regime, where the effects due to the influence of transient
processes have been smoothed out. The time-homogeneous statistical
properties are obtained in the long-time limit after the temporal
and ensemble averaging are performed.

The efficiency (quantum yield) of the proton pumping device  is defined by the formula:
\begin{eqnarray}
\Phi= {\rm \frac{ number \;of\; protons\; pumped}{number\; of\; photons \;absorbed}} \;. \nonumber
\end{eqnarray}
\noindent The photon absorption rate, $\kappa_{\rm{B\rightarrow C}}$, is approximately equal to the rate of light-induced
transitions from the state B to the state C. Thus we assume,
\begin{eqnarray}
\Phi \simeq  \frac{I_{\rm{p}}}{\kappa_{\rm{B\rightarrow C}}}, \label{Eff}
\end{eqnarray}
where $I_{\rm{p}}$ is the proton current (the number of protons, $N_{\rm p},$ translocated across the membrane per unit of time).

\subsection{Diffusive motion of the shuttle in the lipid bilayer}

In Fig.~5 we present the diffusive motion (see Fig.~5(a)) of the
shuttle in  the lipid bilayer together with the time dependencies
of the electron and proton populations of the shuttle (Fig.~5(b))
\begin{figure}[htp]
\centering\includegraphics[width=8.5cm,angle=0,clip]{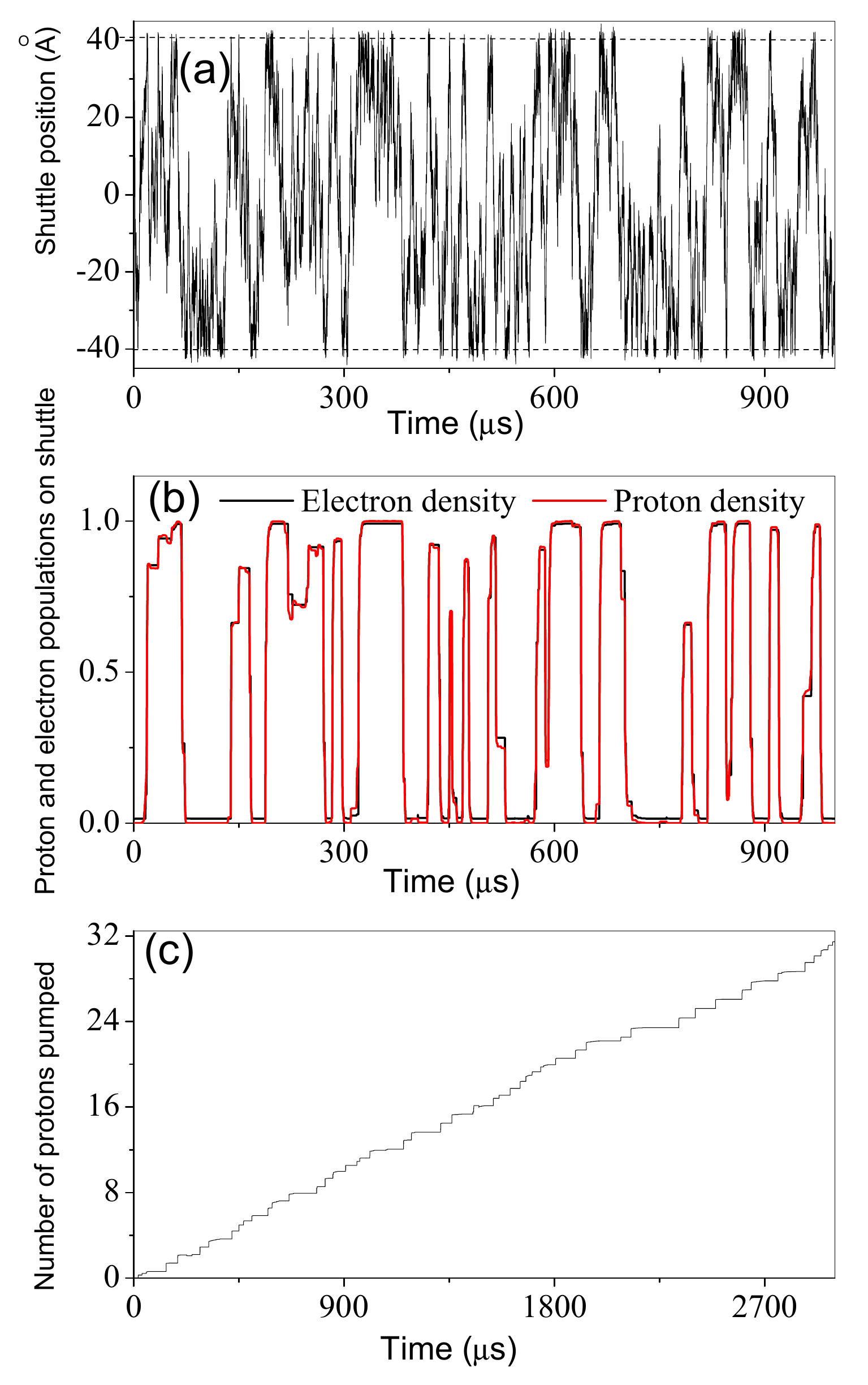}
\caption[]{(Color online) (a) Stochastic motion of the shuttle with
time.  The horizontal black dashed lines denote the borders of the
membrane, $x_{\rm N}$ = 40~\AA, $x_{\rm P}$ = -- 40~\AA.
 Via this diffusion the shuttle transports protons and electrons through the membrane.
 (b) Variation of the electron and proton population on the shuttle. Note that the proton density (red curve) and the electron
 density (black curve) mostly coincide in (b). (c) Number of protons pumped versus time.
The main parameters used here are the light intensity $I=0.138$
mWcm$^{-2}$, temperature $T = 298$ K, and the chemical potentials
$\mu_{\rm{P}}=110$ meV and $\mu_{\rm{N}}=-110$ meV. The light
intensity $I$ corresponds to the photosensitive BC-group with a
dipole moment $\sim |e|\times 1$~nm, where $e$ is the electron
charge.}
\end{figure}
 complemented by the time evolution of the number of
pumped protons (Fig.~5(c)). We assume that the reorganization
energies  for the thermal electron transfers are low enough to
provide a high performance of the system:
\begin{eqnarray}
 \lambda \sim \lambda_{\rm DB} \sim \lambda_{\rm AC}
\sim \lambda_{\rm AS} \sim \lambda_{\rm DS} \sim 400 \;{\rm meV.}\nonumber
\end{eqnarray}
This value of $\lambda$ is quite common for porphyrin-quinone
dyads having a lower limit (the internal reorganization energy) of
the order of 0.3 eV \cite{Heitele94}. Even smaller reorganization
energies ($\lambda \sim$ 230 meV) have been measured for the
porphyrin-fullerene dyads \cite{ImahoriJPC01}. The initial stages
of electron transfer in bacterial reaction centers \cite{Parson98}
are also characterized by a low reorganization energy: $\lambda
\sim$ 70--300 meV, depending on the environment. This is
 due to the fact that the bacteriochlorophyll molecules (and the molecules of porphyrin involved in our molecular triad)
contain highly delocalized $\pi$-electron  systems. In the next section, we also analyze how sensitive the results are to changes in
the values of $\lambda$.

Electrochemical measurements \cite{GustSc89} show that  the energy
of the carotene(D)-porphyrin(BC)-quinone(A) molecular triad sweeps
from the value $\sim$1.9 eV (the first excited state of the
porphyrin, D--B$^1$C--A), to the energy $\sim$1.4 eV, related to
the intermediate state D--BC$^{+}$--A$^{-}$, and, finally, to the
energy, $\sim$1.1 eV, of the charge-separated state
D$^{+}$--BC--A$^{-}$. We assume here that the energy of the first
excited state of the porphyrin, $E_{\rm C} - E_{\rm B}$, is
1908~meV, which corresponds to a photon wavelength of 650 nm as
used in experiments \cite{gali1,gali2}. We have taken the energy
gap between the site C and A to be approximately equal to the
reorganization energy, $ (E_{\rm{C}}-E_{\rm{A}})\sim \lambda = 400$
meV. This gap is about the energy difference between the
D--B$^1$C--A and D--BC$^{+}$--A$^{-}$ states.

The energies of the electron sites S and A are comparable,
$(E_{\rm{A}}-E_{\rm{S}})\simeq 300$ meV, due to a structural
similarity of the quinone shuttle (S) and quinone moiety of the
molecular triad. The protonation of the shuttle leads to the
lowering of the electron energy on site S due to the
electron-proton Coulomb attraction \cite{gali1}, $u_{\rm{SQ}} \sim
360 $ meV. The other Coulomb interaction terms are chosen as
$u_{\rm{DB}}=u_{\rm{BA}}=120$ meV and $u_{\rm{DA}}=60$ meV. These
values correspond to the electrostatic interaction of two charges
located at distances 4 nm and 8 nm, respectively (in a medium with
a dielectric constant $\sim$ 3). Furthermore, we assume that
$E_{\rm{D}}-E_{\rm{B}} = 400$ meV and $\epsilon_{\rm{Q}} =200$~meV.
We have chosen $\epsilon_{\rm{Q}}$ such that, for the above
mentioned parameters, the device works well at the transmembrane
potential difference $\sim$ 200 mV.

We choose $\mu_{\rm{P}} = 110$ meV, $\mu_{\rm{N}}=-110$ meV,  the
resonant tunneling rates $\Delta/\hbar =15\;$ns$^{-1}$, $\Gamma
/\hbar =1.5\;$ns$^{-1}$,  and the reorganization energy  for the
light-induced electron transfer, $\lambda_{\rm{BC}}\sim 80$ meV.
The majority of parameters in our model are deduced from
experimental data. The rates of electron transfer reactions are
given by
\begin{eqnarray}
\kappa_{C \rightarrow A} &\simeq& \kappa_{D \rightarrow B} \simeq 26 \;\mu{\rm s}^{-1} ,\nonumber \\
\kappa_{A \rightarrow S} &\simeq& \kappa_{S \rightarrow A} \simeq  20 \;\mu{\rm s}^{-1} .\nonumber
\end{eqnarray}
Therefore, the loading and unloading time scales of the shuttle are about $0.05\;\mu$s. The shuttle has enough time to be loaded and
unloaded with electrons and protons when it enters the loading/unloading domain with a size about the electron tunneling length,
$L_{\rm tun}\sim$0.5 nm, and the proton transition length, $L_{\rm Q}\sim$0.2 nm. Figures~5(a,b) show a time synchronization between
the spatial motion of the shuttle and the time variations of the shuttle populations.

It follows from Fig.~5(c) that in 1 ms the shuttle performs near 16 trips and translocates $~10$ protons through the membrane,
provided that the light intensity $I$ = 0.133 mWcm$^{-2}$. We assume that the diffusion coefficient $D_{\rm s}$ is of the order of
$2~$nm$^2 \mu {\rm s}^{-1}$ \cite{geyer1}, and the dipole moment of the BC moiety is about $|e|\times$1~nm, where $e$ is the
electron charge. The number of photons absorbed in 1 ms is $\sim \;18$. Thus, the approximate quantum yield $\Phi$ of the pumping
process is $\sim 55 \; \%$. For this parameters, the diffusive motion of the shuttle is the slow and rate-limiting step of the
pumping process.

\subsection{Robustness of the model}
To show a tolerance of the system to variations  of parameters we
explore here the parameter space of our model. Keeping fixed the
\begin{figure}[htp]
\centering\includegraphics[width=8cm,angle=0,clip]{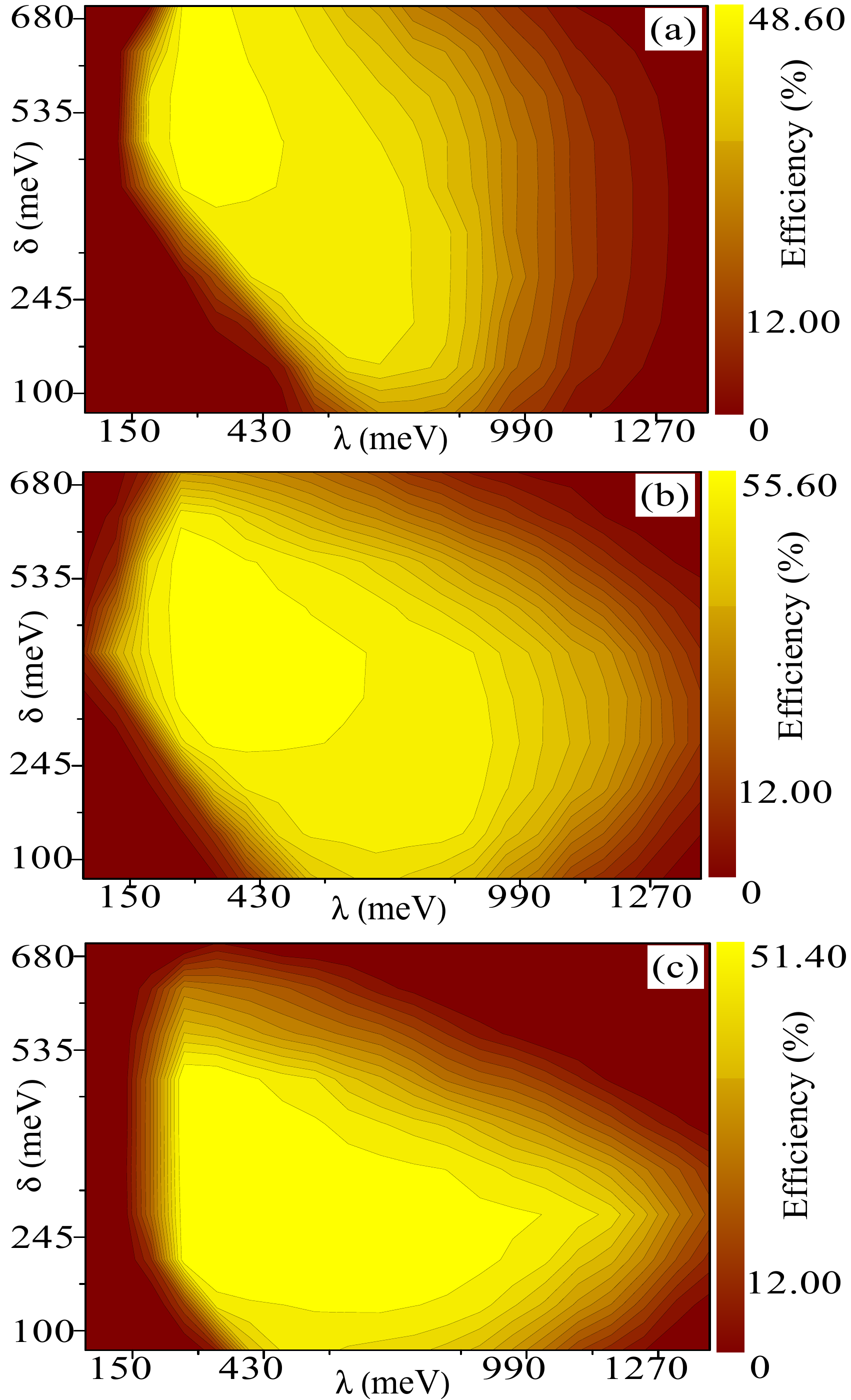}
\caption[]{(Color online) Contour plots presenting the variations
of the quantum efficiency $\Phi$ with the reorganization energy
$\lambda$ and with the energy gap $\delta$, where $\delta =E_{\rm
{C}} - E_{\rm {A}} = E_{\rm {S}} - E_{\rm {D}}$. The parameters
used here are: light intensity $I=0.138$ mWcm$^{-2}$, temperature
$T = 298$ K, and chemical potentials $\mu_{\rm{P}}=110$ meV and
$\mu_{\rm{N}}=-110$ meV. The detunings take the following values:
(a) $E_{\rm{A}} - E_{\rm{S}} = 100$ meV; (b) $E_{\rm{A}} -
E_{\rm{S}} = 300$ meV; (c) $E_{\rm{A}} - E_{\rm{S}} = 500$ meV.}
\end{figure}
energy difference between the sites B and C, we calculate and plot
(see Fig.~6) the pumping efficiency  $\Phi$ (photon-to-proton
quantum yield) as a function of the reorganization energy,
$\lambda$, and the energy gap, $\delta$,
\begin{eqnarray}
 \lambda \sim \lambda_{\rm DB} \sim \lambda_{\rm AC}
\sim \lambda_{\rm AS} \sim \lambda_{\rm DS}, \nonumber\\
 \delta = E_{\rm{C}} - E_{\rm{A}} = E_{\rm{S}} - E_{\rm{D}}, \nonumber
\end{eqnarray}
\noindent between the energy levels $E_{\rm{C}}$ and $E_{\rm{A}}$, and between the levels $E_{\rm{S}}$ and $E_{\rm{D}}$. The figures
6(a), 6(b), 6(c) correspond to the different values of detuning between the acceptor energy level, $E_{\rm{A}}$, and the electron
energy level on the shuttle, $E_{\rm{S}}$: $E_{\rm{A}} - E_{\rm{S}} = 100$ meV (Fig.~6(a)); $E_{\rm{A}} - E_{\rm{S}} = 300$ meV
(Fig.~6(b)); $E_{\rm{A}} - E_{\rm{S}} = 500$ meV (Fig.~6(c)). These plots clearly demonstrate the existence of quite wide areas in
the plane $\lambda$ -- $\delta$, where the pump performs with maximum efficiency. For the detuning $E_{\rm {A}} - E_{\rm {S}}$ = 100
meV (Fig.~6(a)) the pumping efficiency reaches its maximum, $\Phi \sim 48 \%$, in the region of parameters (in meV): $ 270  <
\lambda < 500$, and $400 < \delta < 700$. In this region, the energy gaps between the redox sites are close to the reorganization
energy, which results in higher site-to-site tunneling rates and, consequently, in a high pumping efficiency.

The higher pumping efficiency, $\Phi\sim 55 \%$,  can be achieved at the detuning $E_{\rm {A}} - E_{\rm {S}}$ = 300 meV (Fig.~6(b)).
In this case the parameter $\delta$ can be tuned in such a way that the energy gaps between all relevant electron sites are equal to
the reorganization energy:
\begin{eqnarray}
 &&(E_{\rm{C}} - E_{\rm{A}}) \sim (E_{\rm{A}} - E_{\rm{S}})
 \nonumber\\
 &&\sim (E_{\rm{S}}
  - u_{\rm{SQ}} - E_{\rm{D}}) \sim (E_{\rm{D}} - E_{\rm{B}})
\sim \lambda.\nonumber
\end{eqnarray}
\noindent  We recall that a shuttle populated with a proton has the electron energy, $E_{\rm{S}} - u_{\rm{SQ}}$, which differs from
the initial value $E_{\rm{S}}$ by the charging energy, $u_{\rm{SQ}} \sim 360$ meV. Summing all the above-mentioned detunings and
taking into account the energy difference, $E_{\rm{C}} - E_{\rm{B}} = 1908$ meV, between the optically-active levels B and C, we
estimate the optimum values of the reorganization energy $\lambda $ and the detuning $\delta$:
\begin{eqnarray}
 \lambda \sim \delta
\sim (E_{\rm{C}} - E_{\rm{B}} - u_{\rm{SQ}})/4 = 387 {\rm\; meV}. \nonumber
\end{eqnarray}
The maximum of the efficiency in Fig.~6(b) is observed at $\delta \sim \lambda \sim 400$ meV, which is very close to our
estimations. For a larger energy gap, $E_{\rm{A}} - E_{\rm{S}} = 500$ meV (see Fig.~6(c)), the proton pumping efficiency $\Phi$
decreases and the region of the optimum parameters shrinks compared to Fig.~6(b).

\subsection{Effects of the resonant tunneling rates}

The fine-tuning of tunneling couplings between active electron
sites is  feasible in some nanostructures. This tuning can be
implemented by changing the site-to-site distance, as well as by
varying the height of the potential barriers (see, e.g.,
\cite{Milliron04}).
\begin{figure}[htp]
\centering\includegraphics[width=8cm,angle=0,clip]{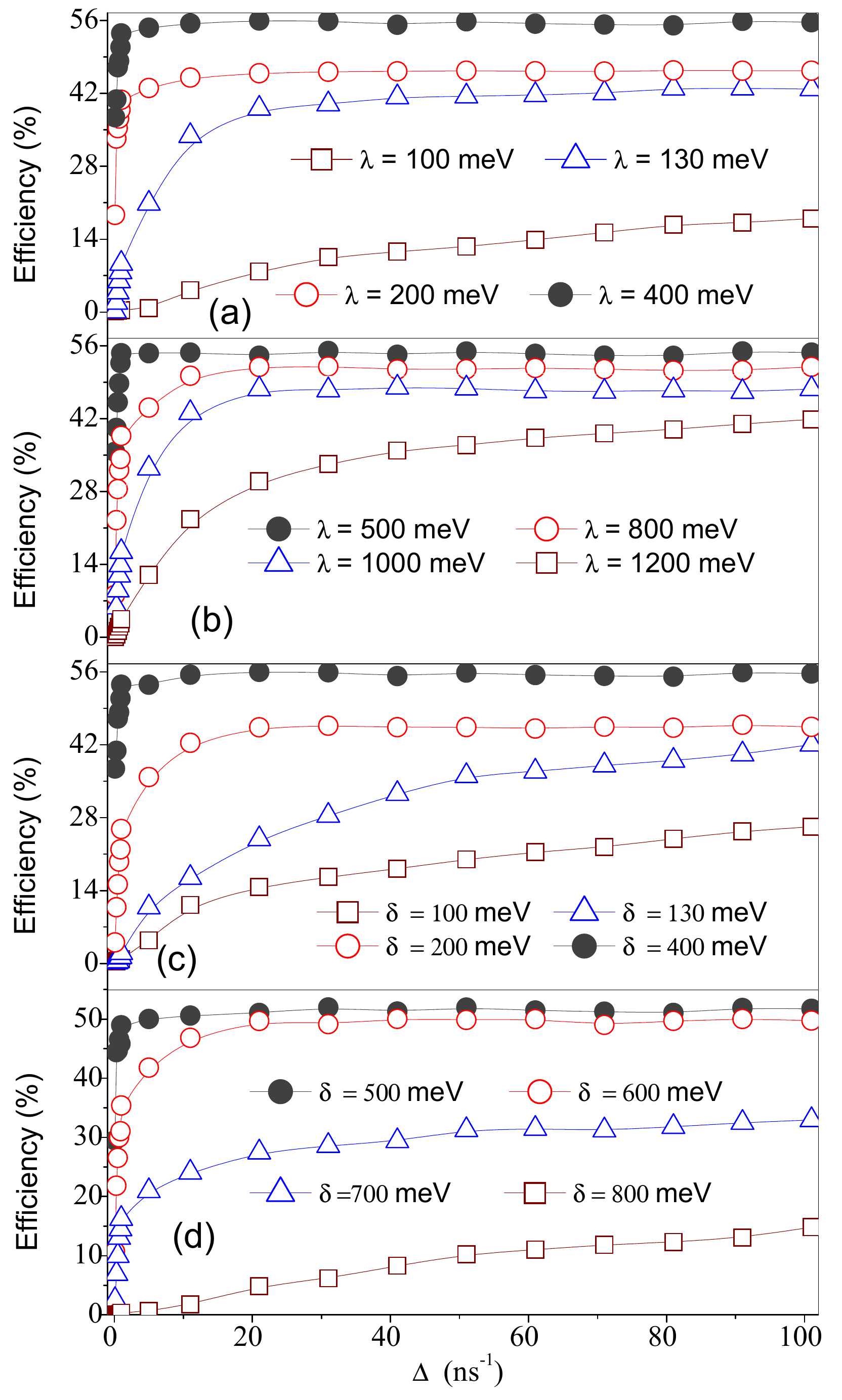}
\caption[]{Color online) Proton pumping quantum efficiency $\Phi$
versus resonant tunneling rate $\Delta$, at different
reorganization energies $\lambda$, shown in  (a,b), and for
different detunings $\delta$, shown in (c,d). Note that $\Delta$
here represents $\Delta/\hbar$, since we set $\hbar = 1.$ We use
the following parameters: $I=0.138$ mWcm$^{-2}$, $T = 298$ K,
$\mu_{\rm{P}}=110$ meV, $\mu_{\rm{N}}=-110$ meV, and the energy gap
$(E_{\rm {A}} - E_{\rm {S}})$ = 300 meV. Panels (a,b) are plotted
at fixed $\delta = 400$ meV, whereas in (c,d) the reorganization
energy is fixed, with $\lambda = 400$ meV.}
\end{figure}
 Artificial photosynthetic systems, such as the
molecular triads, also allow to engineer desirable tunneling and
electrostatic properties of the structures \cite{GustSc89} with the
goal to achieve the highest possible efficiency. As in colloidal
nanocrystals \cite{Milliron04}, this can be done by inserting
additional molecular bridges between the side centers D, A and the
photosensitive part BC utilizing the exponential dependence of
electron tunneling rates on the distance \cite{Wasiel06}.

In Fig.~7 we illustrate the variation of the proton pumping
efficiency $\Phi$ as a function of the resonant tunneling rate
($\Delta/\hbar$) for different values of the reorganization energy
$\lambda$ (Figs.~7(a), 7(b)) and the energy gap $\delta$
(Figs.~7(c), 7(d)). The detuning, $E_{\rm{A}} - E_{\rm{S}}$, is
fixed to the value 300 meV for all plots in Fig.~7.

In Fig.~7(a) we plot four curves, $\Phi(\Delta)$, for the
following set of reorganization energies: $\lambda = 100, \ 130, \
200, \ 400$ meV and for a detuning $\delta = 400$ meV. In Fig.~7(b)
the efficiencies $\Phi(\Delta)$ are plotted for the reorganization
energies: $\lambda = 500, \ 800, \ 1000, \ 1200$ meV, for the same
detuning $\delta$. Similar dependencies, $\Phi(\Delta)$, are
depicted in Fig.~7(c) for $\delta = 100, \ 130, \ 200,\  400$ meV
and in Fig.~7(d) for $\delta = 500,\ 600,\ 700,\  800$ meV. In
both, Figs.~7(c) and 7(d), the reorganization energy $\lambda$ is
equal to 400 meV.

It follows from Fig.~7 that initially the proton pumping
efficiency rapidly increases with increasing $\Delta$, followed by
its saturation for higher values of the resonant tunneling rate.
The saturation limit depends on the reorganization energy $\lambda$
as well as on the energy gap $\delta$. For the optimum values of
$\lambda $ and $\delta$: $\lambda \sim \delta \sim$~400~{\rm meV},
the pumping efficiency is sufficiently high, $\Phi \sim 55$ \%,
even for moderate tunneling rates, $\Delta/\hbar \leq 5$ ns$^{-1}$.

\subsection{Effects of Coulomb interactions}
In Fig.~8 we plot the efficiency $\Phi$ versus the  dielectric
constant $\varepsilon$ of the medium, to explore the effects of the
Coulomb couplings $u_{\rm{DB}}$,  $u_{\rm{BA}}$, and $u_{\rm{DA}}$
on the performance of the proton pump. The electrostatic
\begin{figure}[htp]
\centering\includegraphics[width=8cm,angle=0,clip]{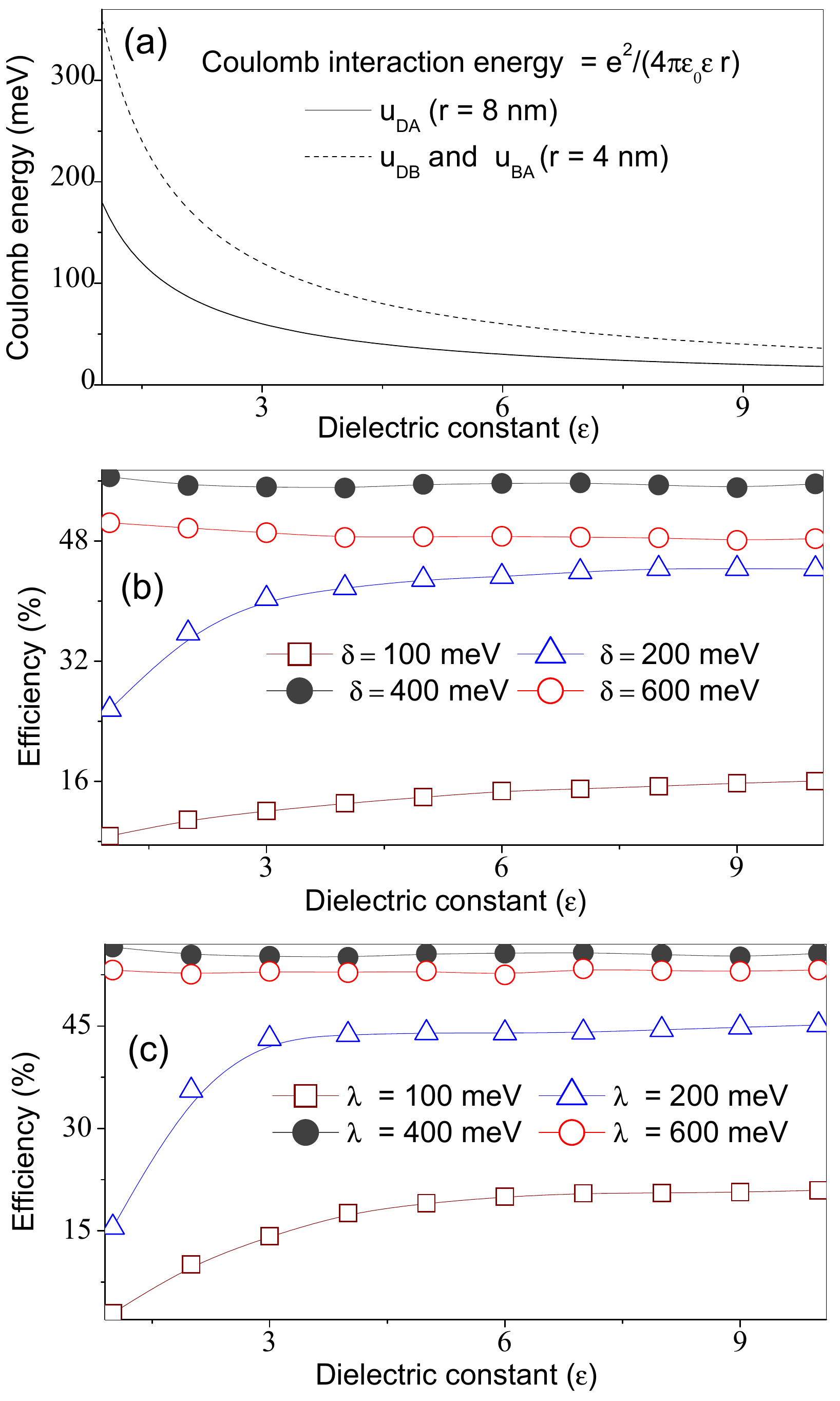}
\caption[]{(Color online) (a) Coulomb energies $u_{\rm{DA}}$ and
$u_{\rm{DB}}$ versus the dielectric constant $\varepsilon$ of the
medium. (b) Proton pumping efficiency $\Phi$ versus dielectric
constant $\varepsilon$ for different values of $\delta$ and for
$\lambda = 400$ meV. (c) The pumping efficiency $\Phi$ as a
function of the dielectric constant $\varepsilon$ for different
reorganization energies $\lambda$ and at the fixed detuning $\delta
= 400$ meV. The other parameters are the same as in Fig.~7:
$I=0.138$ mWcm$^{-2}$, $T = 298$ K, $\mu_{\rm{P}}=110$ meV,
$\mu_{\rm{N}}=-110$ meV, and $E_{\rm {A}} - E_{\rm {S}}$ = 300 meV.
}
\end{figure}
interactions between the photosensitive part B and C and the donor,
$u_{\rm{DB}}$, between the sites B and C and the acceptor,
$u_{\rm{BA}}$, and between the donor and the acceptor,
$u_{\rm{DA}}$, are inversely proportional to the dielectric
constant $\varepsilon$ and to the distance between the relevant
sites. For example, we have
\begin{eqnarray}
u_{DB} = \frac{e^2}{4\pi \,\varepsilon_0\, \varepsilon\, r_{\rm{DB}}}, \nonumber
\end{eqnarray}
\noindent  where $r_{\rm{DB}}$ characterizes the spatial separation of the sites D and B, and $\varepsilon_0$ is the vacuum
permittivity. The Coulomb interactions between the sites D and B and between the sites B and A (with $r_{\rm{DB}} = r_{\rm{BA}}$ = 4
nm) are decreased from 360 meV to 36 meV when the dielectric constant $\varepsilon$ scans the range from 1 to 10 (see Fig.~8(a)). We
note that in our model $r_{\rm{DA}} = 8$ nm, so that $u_{\rm{DA}} = u_{\rm{DB}}/2$. Figure~8(b) shows the efficiencies
$\Phi(\varepsilon)$ for  different values of the detuning $\delta$ : $\delta = 100, \ 200,\ 400,\ 600$ meV, for $\lambda = 400$ meV.
Moreover, in Fig.~8(c) we plot the efficiencies, $\Phi(\varepsilon)$, for fixed detuning $\delta = 400$ meV and for $\lambda = 100,\
200,\ 400,\ 600$ meV.

It should be emphasized that near the optimum working point (at $\lambda \sim \delta \sim 400$ meV) the pump operates with the high
efficiency, $\Phi \sim 55$ \%, which practically does not depend on the dielectric properties of the medium.

\subsection{Effect of light intensity} In Fig.~9 we plot the proton
current as a function of the light intensity for different values of the temperature. At zero light intensity the proton current is
zero. Initially, with increasing light intensity, the proton current also increases linearly and then saturates around $0.2$ mW
cm$^{-2}$. This saturation is probably caused by the slow diffusion of the shuttle inside the lipid membrane. A similar intensity of
saturation ($\sim$ 0.1 mW/cm$^2$) has been observed in experiments \cite{gali2}.
\begin{figure}[htp]
\centering\includegraphics[width=8cm,angle=0,clip]{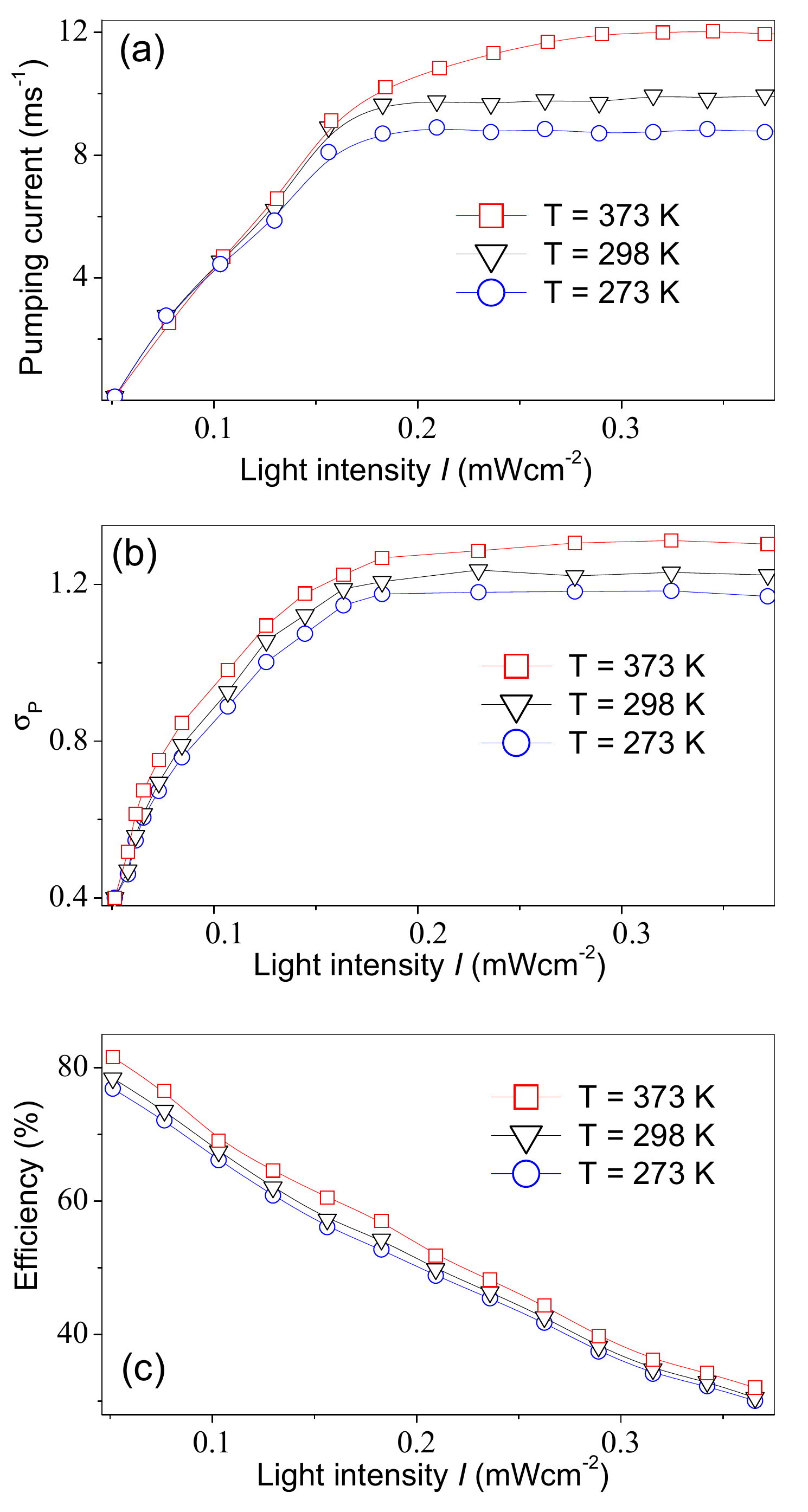}
\caption[]{(Color online) (a)  Proton current versus light
intensity $I$ for different temperatures, at $\mu_{\rm{N}}= - 110$
meV, and $\mu_{\rm{P}}=  110$ meV.
  Notice that the proton current is roughly linear for small intensities
 of light, but it saturates with higher light intensity. In this saturation
 region, the proton current is larger with higher temperatures. (c)
 The standard deviation, $\sigma_{\rm p},$ of the number $N_{\rm p}$ of pumped protons as a function of the light intensity $I$, for
  different temperatures.
 (b) The pumping quantum efficiency $\Phi$ decreases with light intensity for all temperatures shown.
}
\end{figure}

In a warm environment, the shuttle moves faster and carries more protons. To do this, the system should absorb more photons, so that
at high temperatures a full saturation takes place at higher light intensities. We note that the low saturation limit obtained above
and measured in the experiment \cite{gali2} with the carotene-porphyrin-quinone triads is far below the average intensity of solar
light, $I \sim$ 30 mW/cm$^2$ \cite{Hambourger09}. This fact points to the relative inefficiency of the energy-conversion process
available at normal daylight conditions. An ideal highly-efficient photosynthetic system should not have any saturation limits for
the standard daylight intensity of light.

It is evident from Fig.~5 that the number of protons, $N_{\rm p}$, translocated across the membrane fluctuates in time. To estimate
these fluctuations we calculate the standard deviation,
\begin{equation}
\sigma_{\rm p} = \sqrt{\langle N_{\rm p}^2 \rangle - \langle N_{\rm p} \rangle^2}, \label{sigmaP}
\end{equation}
 which characterizes the magnitude of its shot noise. The dependence of the noise level $\sigma_{\rm p}$ on the light
intensity $I$ is shown in Fig.~9(b). For an intensity of light $I \sim$~0.14 mW$^2/{\rm cm}^2$ (when  $\sim$~10 protons are
translocated across the membrane and the efficiency, $\Phi \sim$ 55\%, is sufficiently high) the uncertainty $\sigma_{\rm p}$ in the
number of pumped protons is about 1.3.  In Fig.~9(c) we demonstrate that the efficiency $\Phi$ of the light-induced pumping
decreases monotonically with increasing light intensity. At low light intensities, a relatively small number of photons are absorbed
per unit time. Thus, a higher fraction of the absorbed photons is used for the uphill pumping of the protons.

\subsection{Effect of temperature}
Figure~10 shows the effects of temperature on the pumping current
and on the efficiency of  the photosynthetic device for different
values of the light intensity. The temperature effects appear in
the light-induced proton pumping dynamics through two factors: (i)
The electron transfer rates, including the loading and unloading
rates of the shuttle, increase with increasing temperature. (ii)
The diffusion coefficient of the shuttle increases with
temperature.
\begin{figure}[htp]
\centering\includegraphics[width=8cm,angle=0,clip]{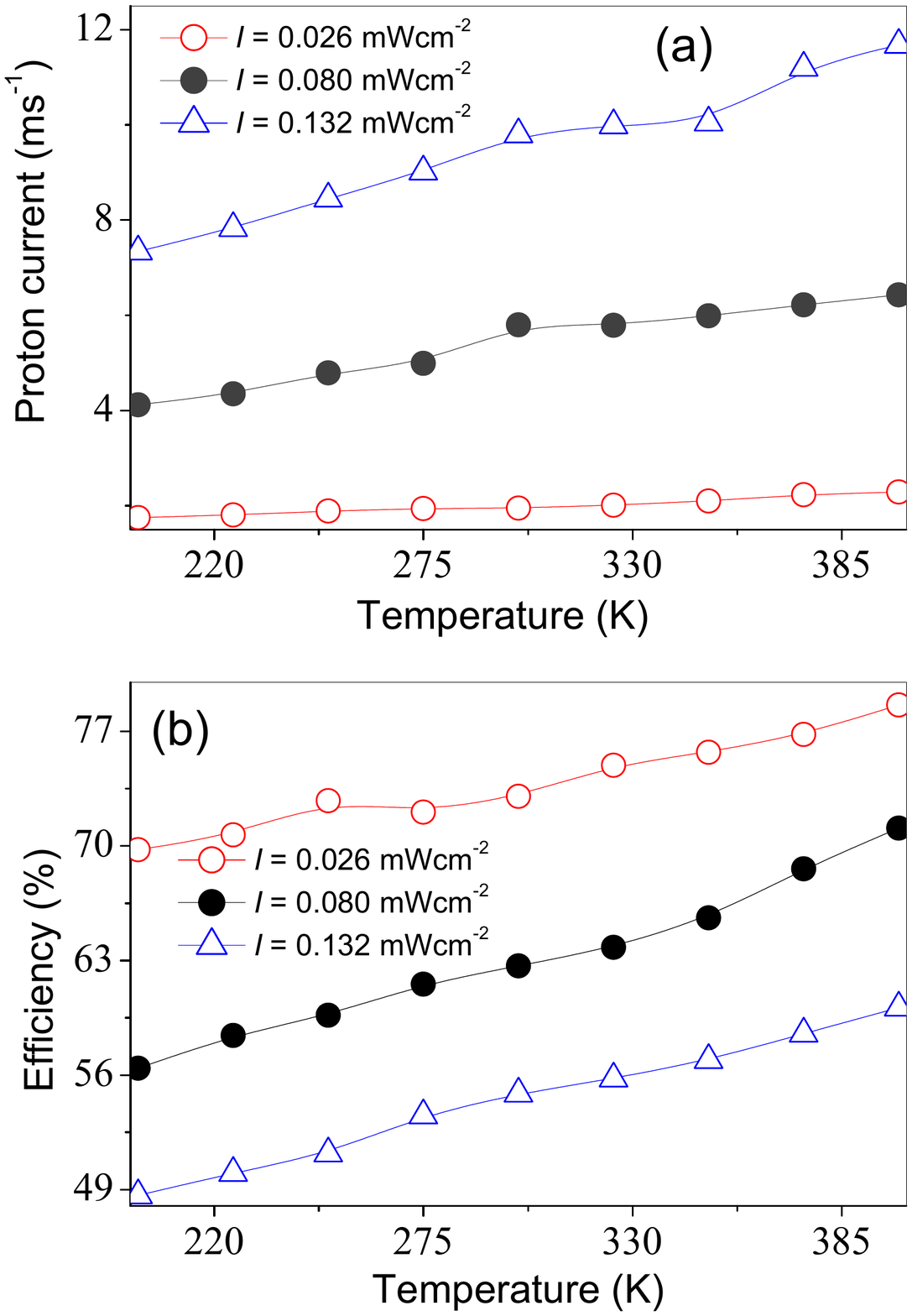}
\caption[]{(Color online)  (a) Proton current versus temperature
for different values of the light intensity $I$. (b) Pumping
efficiency $\Phi$ versus temperature. Here, the electrochemical
gradient $\Delta \mu~=~220 $ meV ($\mu_{\rm{P}}~=~110$ meV, and
$\mu_{\rm{N}}~=~-110$ meV).}
\end{figure}
 Because of this, the shuttle can perform a higher
number of trips to translocate protons at higher temperatures. Here
the electron transfer reactions are not rate-limiting ones. The
diffusive trips of the shuttle from the N terminal to the P
terminal dominate the transfer rate. Therefore, the increase of the
efficiency and the pumping current with temperature is due to the
increase of the number of diffusive trips of the shuttle. A
temperature increase from 200 K to 400 K results in an increase of
about a factor of two in the diffusion constant.
 It is expected that the proton current should increase at the same rate. However, our calculated ratio is about 1.5 (see Fig.~10(a)).
This is, probably, due to  the fact that at high temperatures the shuttle has not enough  time to be completely loaded with
electrons and protons near the acceptor site A and the N-side of the membrane (and unloaded near the donor site D and the P-side of
the membrane). A similar enhancement of the pumping current and the efficiency with temperature
 can be useful for photosynthetic microorganisms to compensate a leakage of protons caused by the high-temperature increase of the
 membrane permeability \cite{Vossenberg95}. The simple physical features which come into play in our model are also important for
 the creation of thermostable artificial photosynthetic devices efficiently converting energy of light into electrical and chemical
 energy in a wide range of temperatures and light intensities.

\subsection{Effect of the electrochemical potential gradient on the proton current}
It follows from Eq.~(\ref{dMu}) that the difference, $\Delta \mu =
\mu_{\rm P} - \mu_{\rm N},$ between the electrochemical potentials
of P- and N-proton reservoirs can be changed by changing the $pH$
levels of the solutions inside and outside of the liposome.
\begin{figure}[htp]
\centering\includegraphics[width=8cm,angle=0,clip]{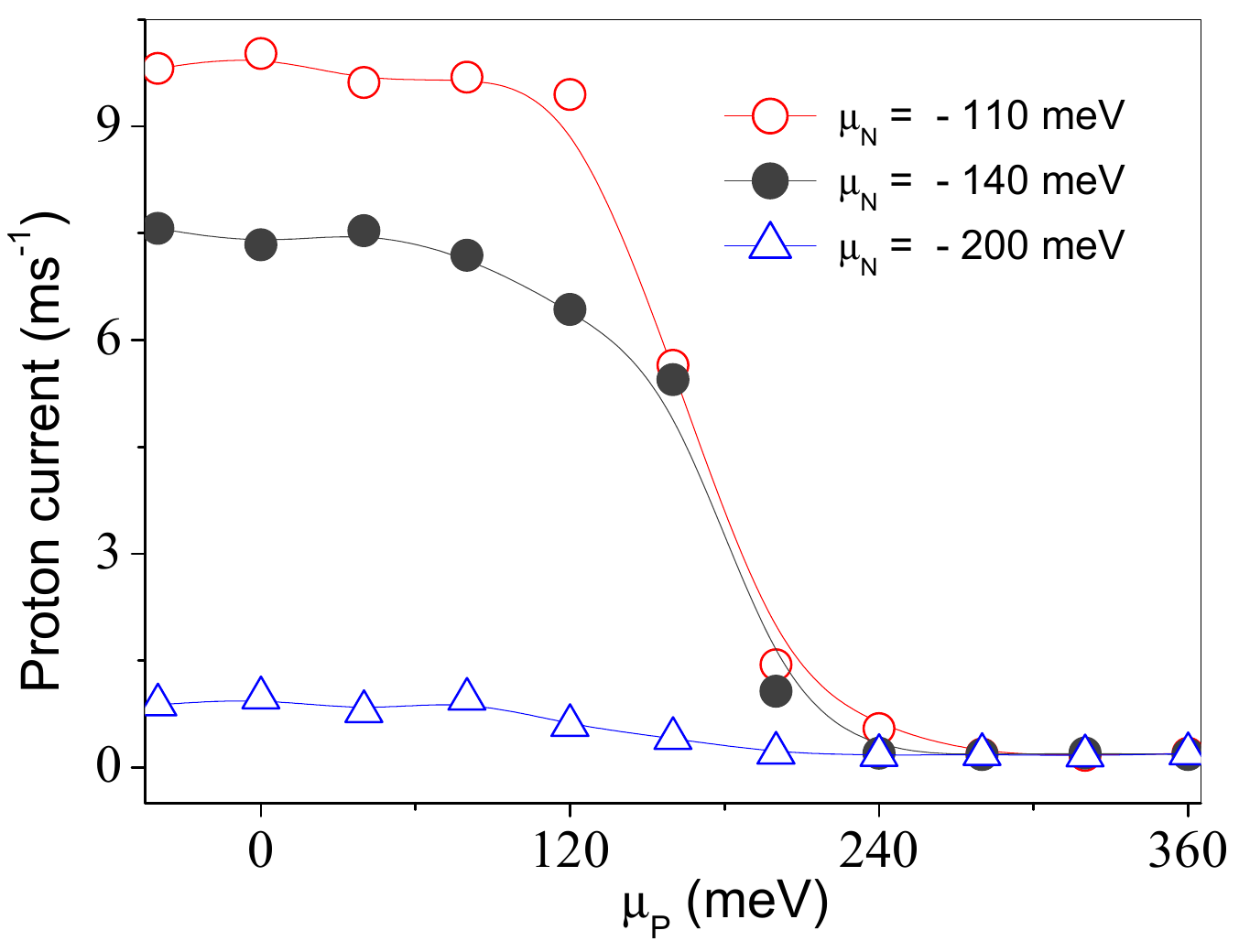}
\caption[]{(Color online) Proton pumping current versus
electrochemical potential $\mu_{\rm{P}}$ of the positive side
(P-reservoir) of the membrane for different values of the potential
$\mu_{\rm{N}}$ of the negative side (N-reservoir) for the light
intensity $I=0.132$ mWcm$^{-2}$ and temperature $T=298$ K.}
\end{figure}
In
doing so, one unit change in $pH$ corresponds to $\sim 59$ meV
variation  of the transmembrane proton gradient $\Delta \mu$ (at
standard conditions). To demonstrate the effect of the $pH$ levels
on the performance of the pump, in Fig.~11 we plot the dependencies
of the proton current on the electrochemical potential $\mu_{\rm
P}$ of the positive side of the membrane at three different values
of the N-side potential: $\mu_{\rm N}$ = --110; --140; and --200
meV. The proton current saturates when the P-side potential is
sufficiently low, $\mu_{\rm P} < 160$~meV, and goes to zero at
$\mu_{\rm P} > 200$~meV). At this condition, the potential of the
P-side exceeds the energy, $\epsilon_Q = 200$~meV, of the proton on
the shuttle: $\mu_{\rm P} > \epsilon_Q$, so that the proton cannot
be translocated to the P-reservoir. On the other hand, the shuttle
cannot be loaded with a proton at the N-side of the membrane if the
electrochemical potential $\mu_{\rm N}$ is below the energy,
$\epsilon_Q - u_{SQ} = - 160$~meV, of the proton on the shuttle
populated with a single electron: $\mu_{\rm N} < -160$~meV. This is
the reason why the last curve in Fig.~11 (taken at $\mu_{\rm N}$ =
--200 meV) goes far below the other two curves (plotted for
$\mu_{\rm N} > -160$~meV).

\section{Conclusions}
We have analyzed a simple model  for light-induced proton pumps in artificial photosynthetic systems. This model has five electron
sites [four sites (D,B,C,A) for the triad molecule and one site for the shuttle (S)] and one proton-binding site on the shuttle (Q).
The shuttle exhibits diffusive motion in the lipid bilayer, so that the electron and proton populations of the shuttle depend on the
shuttle position. Based on the methods of quantum transport theory we have derived and solved numerically a system of master
equations for electron and proton state probabilities evolving in time together with the Langevin equation for the position of the
shuttle. This allows us to calculate the proton current and the pumping efficiency of the system and determine their dependence on:
the intensity of light, temperature and electrochemical potential gradient.

For a reasonable set of parameters, closely related to the experimental setup, we demonstrate that this photosynthetic device can
translocate protons against an electrochemical gradient of the order of 220 meV with the efficiency (photon-to-proton quantum yield)
which exceeds 55\%. Our results explain experiments on artificial photosynthetic reaction centers \cite{gali2}. We predict that both
the proton current and the pumping efficiency grow linearly with temperature due to the related increase of the number of diffusive
trips of the shuttle. We also show that the pumping current increases linearly with the light intensity and saturates at the
experimentally-observed limit, which is lower than the average intensity of solar light.

\section*{Acknowledgments}
We acknowledge partial support from the National Security Agency
(NSA), Laboratory for Physical Sciences (LPS), Army Research Office
(ARO), and National Science Foundation (NSF) grant No. 0726906. We
also acknowledge the RIKEN Super Combined Cluster System for
computational facilities.

\appendix
\begin{appendix}
\section{Basis sets: electron-proton eigenstates and energy
eigenvalues}


The electron-proton system studied here with no leads can be characterized by the 20 basis states of the Hamiltonian $H_0$
\begin{eqnarray}
|1\rangle &=& a^{\dag}_{\rm{D}} a^{\dag}_{\rm{B}} |0\rangle \;;\;\;\;\;\;\; |11\rangle = a^{\dag}_{\rm{D}} a^{\dag}_{\rm{B}}
b^{\dag}_{\rm{Q}} |0\rangle \nonumber
\\
|2\rangle &=& a^{\dag}_{\rm{D}} a^{\dag}_{\rm{C}} |0\rangle \;;\;\;\;\;\;\;|12\rangle = a^{\dag}_{\rm{D}} a^{\dag}_{\rm{C}}
b^{\dag}_{\rm{Q}} |0\rangle \nonumber
\\
|3\rangle  &=& a^{\dag}_{\rm{D}} a^{\dag}_{\rm{A}} |0\rangle \;;\;\;\;\;\;\;|13\rangle = a^{\dag}_{\rm{D}} a^{\dag}_{\rm{A}}
b^{\dag}_{\rm{Q}} |0\rangle \nonumber
\\
|4\rangle  &=& a^{\dag}_{\rm{B}} a^{\dag}_{\rm{C}} |0\rangle \;;\;\;\;\;\;\;|14\rangle = a^{\dag}_{\rm{B}} a^{\dag}_{\rm{C}}
b^{\dag}_{\rm{Q}} |0\rangle \nonumber
\\
|5\rangle &=&  a^{\dag}_{\rm{B}} a^{\dag}_{\rm{A}} |0\rangle \;;\;\;\;\;\;\;|15\rangle = a^{\dag}_{\rm{B}} a^{\dag}_{\rm{A}}
b^{\dag}_{\rm{Q}} |0\rangle \nonumber
\\
|6\rangle  &=& a^{\dag}_{\rm{C}} a^{\dag}_{\rm{A}} |0\rangle \;;\;\;\;\;\;\;|16\rangle = a^{\dag}_{\rm{C}} a^{\dag}_{\rm{A}}
b^{\dag}_{\rm{Q}} |0\rangle \nonumber
\\
|7\rangle  &=& a^{\dag}_{\rm{D}} a^{\dag}_{\rm{S}} |0\rangle \;;\;\;\;\;\;\;|17\rangle = a^{\dag}_{\rm{D}} a^{\dag}_{\rm{S}}
b^{\dag}_{\rm{Q}} |0\rangle \nonumber
\\
|8\rangle  &=& a^{\dag}_{\rm{B}} a^{\dag}_{\rm{S}} |0\rangle \;;\;\;\;\;\;\;|18\rangle = a^{\dag}_{\rm{B}} a^{\dag}_{\rm{S}}
b^{\dag}_{\rm{Q}} |0\rangle \nonumber
\\|9\rangle &=& a^{\dag}_{\rm{C}} a^{\dag}_{\rm{S}} |0\rangle
\;;\;\;\;\;\;\;|19\rangle = a^{\dag}_{\rm{C}} a^{\dag}_{\rm{S}} b^{\dag}_{\rm{Q}} |0\rangle \nonumber
\\|10\rangle &=& a^{\dag}_{\rm{A}} a^{\dag}_{\rm{S}} |0\rangle
\;;\;\;\;\;\;\;|20\rangle = a^{\dag}_{\rm{A}} a^{\dag}_{\rm{S}} b^{\dag}_{\rm{Q}} |0\rangle
\end{eqnarray}
Here, $|0\rangle$  represents the vacuum state, when all electron and proton sites are empty. The state $|1\rangle =
a^{\dag}_{\rm{D}} a^{\dag}_{\rm{B}} |0\rangle$ corresponds to the case when one electron is located on the site D and one on the
site B, and so on. The state $|11\rangle = a^{\dag}_{\rm{D}} a^{\dag}_{\rm{B}} b^{\dag}_{\rm{Q}} |0\rangle $ indicates that, in
addition to two electrons on the sites D and B, there is also a proton on the shuttle. The states $|1\rangle $ to $|10\rangle$
describe the shuttle with no protons, whereas the states $|11\rangle $ to $|20\rangle$ are related to the shuttle populated with a
single proton.

An arbitrary operator $A$ of the combined   electron-proton system can be expressed in terms of the basis Heisenberg matrices
$\rho_{m,n}=|m\rangle \langle n|$:
$$A=\sum_{m,n} A_{mn}\;\rho_{m,n},$$
where $m$ and $n$ label the basis states:  $m,n = 1,\ldots,20.$ The diagonal operator is denoted as: $\rho_m \equiv \rho_{m,m}$.
Thus the electron population operators $\{n_{\rm{D}}, n_{\rm{B}}, n_{\rm{C}}, n_{\rm{A}}, n_{\rm{S}}\}$ can be represented in the
form
\begin{eqnarray}
n_{\rm{D}} &=&\rho_{1}+\rho_{2}+\rho_{3}+\rho_{7}+\rho_{11} +\rho_{12}+\rho_{13}+\rho_{17}\nonumber
\\
n_{\rm{B}} &=&\rho_{1}+\rho_{4}+\rho_{5}+\rho_{8}+\rho_{11} +\rho_{14}+\rho_{15}+\rho_{18}\nonumber
\\
n_{\rm{C}} &=&\rho_{2}+\rho_{4}+\rho_{6}+\rho_{9}+\rho_{12} +\rho_{14}+\rho_{16}+\rho_{19}\nonumber
\\
n_{\rm{A}} &=&\rho_{3}+\rho_{5}+\rho_{6}+\rho_{10}+\rho_{13} +\rho_{15}+\rho_{16}+\rho_{20}\nonumber
\\
n_{\rm{S}} &=&\rho_{7}+\rho_{8}+\rho_{9}+\rho_{10}+\rho_{17}
+\rho_{18}+\rho_{19}+\rho_{20},\nonumber\\
\end{eqnarray}
and for the operator of the proton population of the shuttle we obtain
\begin{equation}
 n_{\rm{Q}} =\rho_{11}+\rho_{12}+\rho_{13}+\rho_{14}+\rho_{15}
 +\rho_{16}+\rho_{17} +\rho_{18}+\rho_{19}+\rho_{20}.
\end{equation}
 Using the eigenfunctions (see Eq.~(A1)), we can rewrite the Hamiltonian
 $H_0$ in a simple diagonal form:
\begin{eqnarray}
H_0 = \sum_{m=1}^{20} \varepsilon_{m} \rho_m.
\end{eqnarray}
with the following energy spectrum:
\begin{eqnarray}
\varepsilon_1 &=& E_{\rm{D}}+E_{\rm{B}}\;;\;\;\;\;\;\;\;\;\varepsilon_{11}= \varepsilon_1 + \epsilon_{\rm{Q}} \nonumber
\\
\varepsilon_2 &=& E_{\rm{D}}+E_{\rm{C}} \;;\;\;\;\;\;\;\;\;\varepsilon_{12}= \varepsilon_2 + \epsilon_{\rm{Q}} \nonumber
\\
\varepsilon_3 &=& E_{\rm{D}}+E_{\rm{A}}-u_{\rm{BA}}
\;;\;\;\;\;\;\;\; \;\varepsilon_{13}= \varepsilon_3 +
\epsilon_{\rm{Q}} \nonumber
\\
\varepsilon_4 &=& E_{\rm{B}}+E_{\rm{C}}-u_{\rm{DB}}
\;;\;\;\;\;\;\;\;\;\varepsilon_{14}= \varepsilon_4 +
\epsilon_{\rm{Q}} \nonumber
\\
\varepsilon_5 &=& E_{\rm{B}}+E_{\rm{A}}-u_{\rm{DA}}
\;;\;\;\;\;\;\;\;\;\varepsilon_{15}= \varepsilon_5 +
\epsilon_{\rm{Q}} \nonumber
\\
\varepsilon_6 &=& E_{\rm{C}}+E_{\rm{A}}-u_{\rm{DA}} \;;\;\;\;\;\;\;\;\;\varepsilon_{16}= \varepsilon_6 + \epsilon_{\rm{Q}} \nonumber
\end{eqnarray}
\begin{eqnarray}
\varepsilon_7 &=& E_{\rm{D}}+E_{\rm{S}}
\;;\;\;\;\;\;\;\;\;\varepsilon_{17}= \varepsilon_7 +
\epsilon_{\rm{Q}}-u_{\rm{SQ}} \nonumber
\\
\varepsilon_8 &=& E_{\rm{B}}+E_{\rm{S}} \;;\;\;\;\;\;\;\;\;\varepsilon_{18}= \varepsilon_8 + \epsilon_{\rm{Q}}-u_{\rm{SQ}} \nonumber
\\
\varepsilon_9 &=& E_{\rm{C}}+E_{\rm{S}}
\;;\;\;\;\;\;\;\;\;\varepsilon_{19}= \varepsilon_9 +
\epsilon_{\rm{Q}}-u_{\rm{SQ}} \nonumber
\\
\varepsilon_{10} &=&
E_{\rm{A}}+E_{\rm{S}}+u_{\rm{DB}}-u_{\rm{BA}}-u_{\rm{DA}}\nonumber
\\
\varepsilon_{20} &=& \varepsilon_{10} +
\epsilon_{\rm{Q}}-u_{\rm{SQ}}.
\end{eqnarray}
The terms $a^{\dag}_i a_{i'}$, describing the direct tunneling between all possible coupled sites $i$ and $i'$ are given by the
expressions
\begin{eqnarray}
a^{\dag}_{\rm{B}} a_{\rm{D}} &=& \rho_{4,2}+\rho_{5,3}+\rho_{8,7}+\rho_{14,12}+\rho_{15,13}+\rho_{18,17}\nonumber
\\
a^{\dag}_{\rm{A}} a_{\rm{C}} &=& \rho_{3,2}+\rho_{5,4}+\rho_{10,9}+\rho_{13,12}+\rho_{15,14}+\rho_{20,19}\nonumber
\\
a^{\dag}_{\rm{B}} a_{\rm{C}} &=& \rho_{1,2}+\rho_{5,6}+\rho_{8,9}+\rho_{11,12}+\rho_{15,16}+\rho_{18,19}\nonumber
\\
a^{\dag}_{\rm{A}} a_{\rm{S}} &=& \rho_{3,7}+\rho_{5,8}+\rho_{6,9}+\rho_{13,17}+\rho_{15,18}+\rho_{16,19}\nonumber
\\
a^{\dag}_{\rm{D}} a_{\rm{S}} &=&
-\rho_{1,8}-\rho_{2,9}-\rho_{3,10}-
\rho_{11,18}-\rho_{12,19}-\rho_{13,20}\nonumber\\
\end{eqnarray}
It should be noted that the operator $H_{\rm{dir}}$  in Eq.~(3) is non-diagonal. The proton operator $b_{\rm{Q}}$ can also be
expressed in a similar form
\begin{equation}
b_{\rm{Q}} = \sum_{mn} b_{Q,mn} \rho_{m,n}.
\end{equation}
\end{appendix}

\end{document}